\documentclass[aps,prb,twocolumn,superscriptaddress,floatfix]{revtex4-1}
\usepackage{amsfonts,bm}
\usepackage{amsmath}
\usepackage{amssymb}
\usepackage{amsthm}
\usepackage{graphicx}
\usepackage{color}
\usepackage{bm}
\usepackage{hyperref}
\usepackage{rotating}
\usepackage{comment}
\usepackage{indentfirst}
\usepackage[normalem]{ulem}
\usepackage{epstopdf}
\usepackage{ragged2e}
\hypersetup{
    colorlinks,
    citecolor=red,
    filecolor=black,
    linkcolor=black,
    urlcolor=black
}
\pdfoutput=1

\newcommand{\<}{\langle}
\renewcommand{\>}{\rangle}
\newcommand{\be}{\begin{equation} }
\newcommand{\ee}{\end{equation} }
\newcommand{\ba}{\begin{eqnarray} }
\newcommand{\ea}{\end{eqnarray} }
\newcommand{\n}{\nonumber \\ }

\newcommand{\mac}{\mathcal}

\newcommand{\bpm}{\begin{pmatrix}}
\newcommand{\epm}{\end{pmatrix}}
\newcommand{\bmm}{\begin{matrix}}
\newcommand{\emm}{\end{matrix}}

\newcommand{\bea}{\begin{eqnarray}}
\newcommand{\eea}{\end{eqnarray}}

\newcommand{\beq}{\begin{equation} }
\newcommand{\eeq}{\end{equation} }
\newcommand{\non}{\nonumber }
\newcommand{\Z}{{\mathbb{Z}}}

\newcommand{\mz}{\mathbb{Z}}

\newcommand{\bit}{\begin{itemize}}
\newcommand{\eit}{\end{itemize}}

\begin{document}

%\title{Symmetry enriched string-net models: Exactly solvable models for SET phases}

%\title{Symmetry enriched string-net construction: Exactly solvable models for SET phases}

\title{Symmetry enriched string-nets: Exactly solvable models for SET phases}

 \author{Chris Heinrich}
\affiliation{Department of Physics, James Frank Institute, University of Chicago, Chicago, Illinois 60637,  USA}
\author{Fiona Burnell}
\affiliation{Department of Physics and Astronomy, University of Minnesota, Minneapolis, Minnesota 55455, USA}
\author{Lukasz Fidkowski}
\affiliation{Department of Physics and Astronomy, Stony Brook University, Stony Brook, NY 11794-3800, USA}
\author{Michael Levin}
\affiliation{Department of Physics, James Frank Institute, University of Chicago, Chicago, Illinois 60637,  USA}
%\date{\today}

\begin{abstract}

We construct exactly solvable models for a wide class of symmetry enriched topological (SET) phases. Our construction applies to 2D bosonic SET phases with finite unitary onsite symmetry group $G$ and we conjecture that our models realize every phase in this class that can be described by a commuting projector Hamiltonian. Our models are designed so that they have a special property: if we couple them to a dynamical lattice gauge field with gauge group $G$, the resulting gauge theories \ {are equivalent to string-net models}. This property is what allows us to analyze our models in generality. As an example, we present a model for a phase with the same anyon excitations as the toric code and with a $\mathbb{Z}_2$ symmetry which exchanges the $e$ and $m$ type anyons. We further illustrate our construction with a number of additional examples.

\end{abstract}

\maketitle

%\tableofcontents

%%%%%%%%%%%%%%%%%%%%%%%%%%%%%%%%
%%%%%%%%%%%%%%%%%%%%%%%%%%%%%%%%
\section{Introduction}

The interplay between symmetry and topology in gapped quantum many body systems has been a subject of intense investigation recently. Spurred by the theoretical prediction and experimental discovery of topological insulators,\cite{hasan10, qi11} a large class of gapped quantum many body systems in which symmetry and topology play a crucial role have been predicted and classified. At the heart of this classification is the concept of a gapped \emph{phase}: two gapped many body systems are said to belong to the same phase if one can continuously interpolate between their Hamiltonians while maintaining a finite energy gap and preserving all physical symmetries.

A useful approach for studying gapped quantum phases, especially those with interactions, is to construct exactly solvable lattice models that realize them.\cite{kitaev2003fault, string-net, Kitaev2006, Haah, WalkerWang, chen2013symmetry} Among other applications, such models have been used to prove that certain phases exist and are anomaly free.\cite{Haah, BCFVSPTmodel} In addition, exactly solvable models have revealed new properties of previously known phases (see e.g. Refs. \onlinecite{BravyiKitaev, LevinWenEnt}). In most cases, the solvability of these models comes from the fact that their Hamiltonians can be written as a sum of commuting projection operators $P_i$, i.e. $H = -\sum_i P_i$. We will refer to Hamiltonians of this type as `commuting projector' models.

The solvable model approach has been developed primarily in the context of two types of phases: (1) phases that support anyon excitations but have no symmetries and (2) phases that \emph{don't} support anyon excitations but \emph{do} have symmetries. The first type of phase is commonly called a `topological' phase, while the second type is known as a `symmetry-protected topological' phase. In the former case, the string-net models of Ref. \onlinecite{string-net} are known to realize a large class of two dimensional topological phases. Likewise, in the latter case, the cohomology models of Ref. \onlinecite{chen2013symmetry} can realize symmetry-protected topological (SPT) phases in arbitrary spatial dimension with finite unitary onsite symmetry. These constructions are especially appealing because they are conjectured to realize \emph{all} phases of type (1) and (2) that can be built from commuting projector models. 

In view of these successes, it is natural to try to build solvable models for more general phases that have \emph{both} anyon excitations and symmetry. Such phases are known as symmetry-\emph{enriched} topological (SET) phases. Some progress has been made in this direction and a number of solvable models for SET phases have been written down.\cite{Mesaros_Ran, Hermele1, Teo_Hughes, Tarantino1}
However, these constructions are not as general as they could be since they do not include all phases that can be realized with commuting projector models.

In this paper we construct exactly solvable models for more general 2D SET phases, which we call  `symmetry enriched string-net' models. Our construction applies to the simplest class of 2D SET phases, namely those that are built out of bosonic degrees of freedom and have a finite unitary onsite symmetry. We conjecture that, within this class of SET phases, our models realize every phase that can be described by a commuting projector Hamiltonian. (In fact, our conjecture is even stronger: we believe that our models realize every phase whose underlying topological order can be realized by a commuting projector model.) As an example, we present an exactly solvable model for an SET phase that has eluded previous constructions, namely a phase with the same anyon excitations \ {as the toric code\cite{kitaev2003fault} and} an onsite $\mz_2$ symmetry which exchanges the $e$ and $m$ (i.e. $\mz_2$ charge and $\mz_2$ flux) anyon excitations. 

To understand the basis of our conjecture, we need to recall recent work on the classification of 2D SET phases.\cite{stationq} According to this work, every SET phase with anyon excitations $\mathcal A$ and finite unitary onsite symmetry group $G$ is associated with a mathematical object known as a `braided $G$-crossed extension of $\mathcal{A}$.\cite{stationq,Teo_Hughes} Roughly speaking, this object describes the collective fusion and braiding data of the anyon excitations in $\mathcal A$ along with extrinsic symmetry defects. It is known that if two models are described by distinct braided $G$-crossed extensions then they belong to distinct SET phases. It has also been conjectured that the converse is true.\cite{stationq} If this is the case, then braided $G$-crossed extensions provide a complete classification of 2D SET phases.\footnote{See Ref. \onlinecite{lan2016classification} for a different approach to classification}

Using this language, we can precisely characterize the generality of our models. Consider a topological phase with anyon excitations $\mathcal A$ that is realizable with a string-net model and has no symmetries. 
%[As we will explain later, this is equivalent to the condition that $\mathcal A$ is the Drinfeld center of some fusion category $\mathcal C$, i.e. $A = \mathcal Z(\mathcal C)$]. 
Then, with our construction, we can build a symmetry enriched string-net model realizing every braided $G$-crossed extension 
of $\mathcal A$. Our conjecture now follows immediately from this result if we make two assumptions: (1) string-net models realize every topological phase (without symmetry) that can be realized by commuting projector models and (2) the above classification is correct.

%An interesting corollary of our construction is that all of the braided $G$-crossed extensions of $\mathcal A$ are physically realizable if $\mathcal A$
%can be realized by a string-net model. This is a non-trivial result because in principle there could be physical obstructions to realizing certain 
%$G$-crossed extensions in microscopic lattice models that are not captured by the known mathematical structure. Our construction proves 
%that no such obstructions exist --- at least for a broad class of $G$-crossed extensions.

The general idea behind our construction is as follows. Suppose we want to construct a model for a particular SET phase with symmetry group $G$.
We can build such a model in two steps. The first step is to build a string-net model that realizes the \emph{gauged} SET phase --- that is, the phase obtained by gauging the global $G$-symmetry of the SET phase of interest. In general, this step can be challenging, but mathematical results guarantee that such a string-net model exists. The second step is to `ungauge' the resulting string-net model: that is, to construct a model that has a symmetry group $G$, and with the property that gauging this symmetry gives back the string-net model (or at least something that belongs to the same phase). Conveniently, this second step can be accomplished rather easily by making small modifications to the string-net Hilbert space and Hamiltonian. Once we make these modifications, the resulting lattice model realizes the SET phase of interest.

We now comment on the relationship between our results and previous work. Exactly solvable models realizing certain SET phases were written down in Ref. \onlinecite{Hermele1}. This construction was then extended in Ref. \onlinecite{Tarantino1} to include certain anyon permuting symmetries, but not all. The previous constructions of SET phases that are perhaps most similar to ours are given in Refs. \onlinecite{Mesaros_Ran} and \onlinecite{chang2015enriching}. The models of Ref. \onlinecite{Mesaros_Ran} realize a subset of the SET phases discussed in this paper, and in these cases their models are essentially equivalent to ours. Meanwhile, the models of Ref. \onlinecite{chang2015enriching} are similar to ours in that they realize SET phases by extending the string-net construction. However, our construction has the advantage of providing a more algorithmic way of building models and making the symmetry manifest. It is also worth noting that the 2D group cohomology SPT models\cite{chen2013symmetry} and string-net models\cite{string-net} can be considered as special cases of our construction by taking the limits $\mac{A}=0$ and $G=0$ respectively.

This paper is organized as follows. To warm-up, we present a simple example of our construction in section \ref{toric-code}. 
This example is an exactly solvable model for a toric code SET phase with a $\mz_2$ symmetry which exchanges the $e$ and $m$ anyon 
excitations. In section \ref{relation-to-doubled-ising}, we explain the relationship between this toric code model and the 
string-net model that it descends from, namely the \emph{doubled Ising} string-net model. After this warm-up, we outline the 
general construction in section \ref{general-framework}. We then illustrate the general construction with additional examples 
in section \ref{examples}. Technical arguments are given in the appendices.

\section{Toric code with $e\leftrightarrow m$ symmetry}\label{toric-code}

Before discussing the general construction, it is useful to first see a concrete example. The example that we
present is an exactly solvable model that has the same types of anyon excitations as the toric code\cite{kitaev2003fault} and 
has a global onsite $\mz_2$ symmetry which exchanges the `$e$' and `$m$' type anyons (also known as the $\mz_2$ 
charge and $\mz_2$ flux). We will refer to this example as the `symmetric toric code' model. 
%This example
%is interesting because, despite its simplicity, it has proven challenging to realize with commuting projector
%models.

Like all of the symmetry enriched string-net models, the symmetric toric code model is derived from a parent string-net model --- in
this case, the doubled Ising string-net model. That being said, in this section we will make every effort to 
analyze the symmetric toric code model from first principles, without referring to the 
doubled Ising string-net model. We postpone the discussion of the connection between the two models to 
section \ref{relation-to-doubled-ising}.

We note here that there are actually two \emph{distinct} toric code SET phases with an $e\leftrightarrow m$ symmetry, which differ in a very minor way.\cite{stationq} For simplicity, we focus on just one of them in the main text, but we briefly discuss the other in appendix \ref{other-toric-code}.

%---------------------------------
%---------------------------------
\subsection{The model}
\begin{figure}[tb]
      \centering
    \includegraphics[width=.48\textwidth]{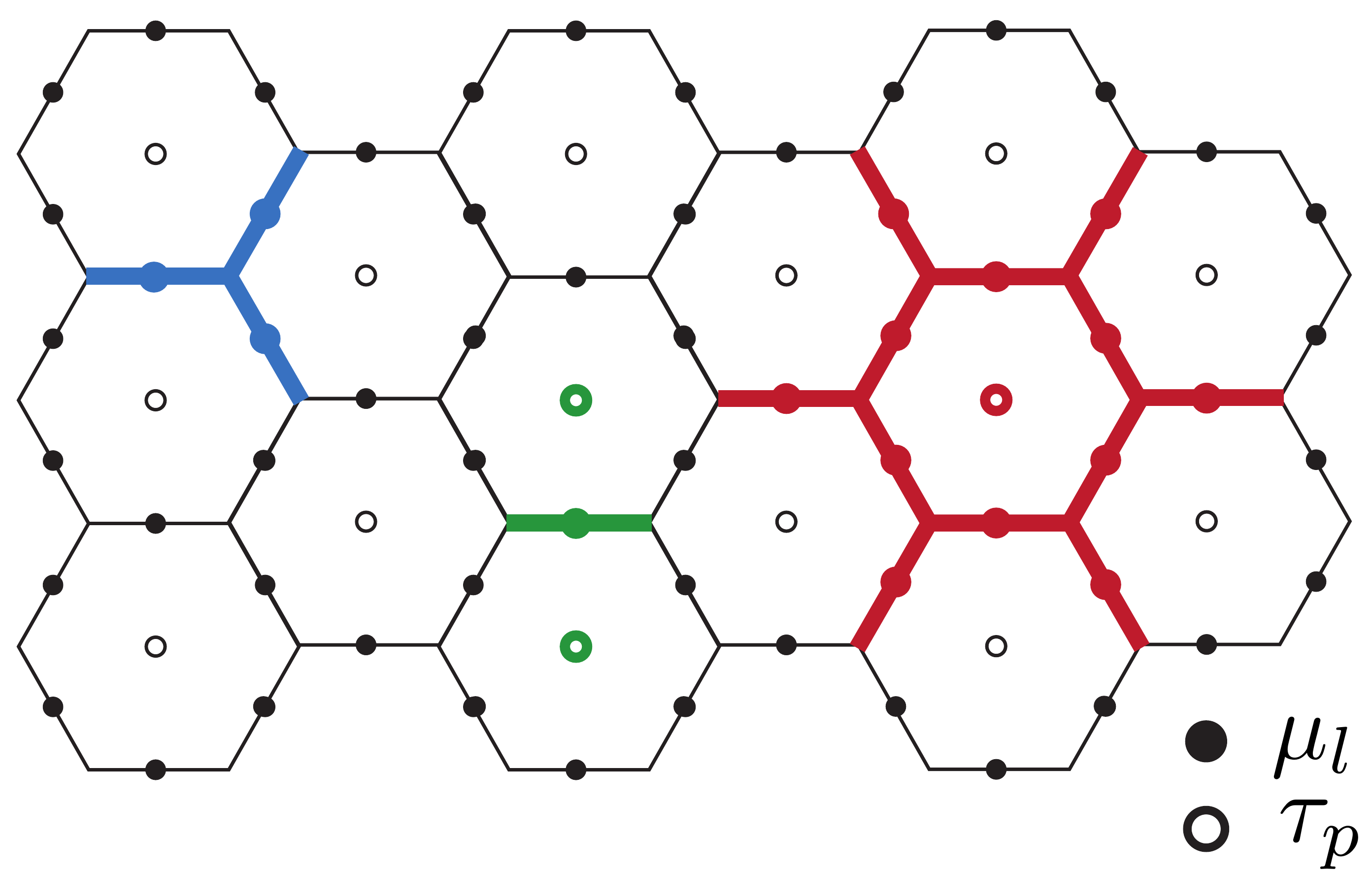}
            \caption{(color online) The Hilbert space of the symmetric toric code model is built out of two-state spins $\tau_p$ and three-state spins $\mu_l$ living on the plaquettes $p$ and links $l$ of the honeycomb lattice. The Hamiltonian is a sum of three terms, 
$P_l, Q_v, B_p$, which act on the green, blue, and red spins, respectively.}
    \label{plaquette-operator}
    \end{figure}

The model is a spin system built out of \emph{two-state} spins that live on the plaquettes of the honeycomb lattice and \emph{three-state} spins that live on the links of the honeycomb lattice. We will denote the basis states of the two-state spins by  $|+\>$ and $|-\>$, and the basis states of the three-state spins by $|1\>$, $|\psi\>$, and $|\sigma\>$. In this notation, the basis states for the full Hilbert space are labeled as
$|\{\tau_p^z, \mu_l\}\>$ where $\tau_p^z= \pm$ parameterizes the states of the plaquette $p$ and $\mu_l = 1,\psi,\sigma$ parameterizes the states of the link $l$ (Fig. \ref{plaquette-operator}).

We will sometimes find it convenient to describe the link degrees of freedom using an alternative language involving strings. In particular, if a link is in the state $|\psi\>$, we will say that it is occupied by a $\psi$ string, and likewise if the link is in the state $|\sigma\>$, we will say it is occupied by a $\sigma$ string. If a link is in the state $|1\>$, we will say that it is unoccupied. Using this language, every state of the link spins $\mu_l$ corresponds to a configuration of $\psi$ and $\sigma$ strings drawn on the honeycomb lattice.

%As we mentioned earlier, a crucial aspect of the model is that it possesses a global $\mz_2$ symmetry. This symmetry is defined by
%\beq
%S=\prod_p\tau^x_p,
%\eeq
%where $\tau^x_p$ denotes the Pauli operator acting on the plaquette spin $p$. In other words, $S$ is a conventional Ising symmetry which flips all the plaquette spins from $|\up\> \leftrightarrow |\down\>$.

The Hamiltonian of the model,
\beq\label{toric-code-ham}
H=-\sum_l P_l -\sum_v Q_v-\sum_p B_p,
\eeq
is expressed as a sum over operators associated with the links ($l$), vertices ($v$), and plaquettes ($p$) of the honeycomb lattice. 
We now explain how each of these operators are defined. 

The link operator $P_l$ acts on three spins --- namely the ones living on the link $l$ and the two adjacent plaquettes $p$ and $q$ (green spins in Fig. \ref{plaquette-operator}). It is defined by
\beq
P_l = \frac{1}{2}(1+(-1)^{N_{l\sigma}} \cdot \tau^z_p \tau^z_q)
\label{link-projector}
\eeq
Here $\tau^z_p$ and $\tau^z_q$ are the standard Pauli operators for the plaquette spins $p$ and $q$, and $N_{l \sigma}$ is a link spin operator defined by $N_{l\sigma}|1\> = N_{l\sigma}|\psi\> = 0$ and $N_{l\sigma} |\sigma\> = |\sigma\>$.
%$N_{l\sigma} = 1$ if $l$ is in the state $|\sigma\>$ and $N_{l\sigma} = 0$ if $l$ is in the state $|1\>$ or $|\psi\>$.

If we examine the above definition we can see that $P_l$ has a simple interpretation: $P_l$ projects onto states such that either (1) $\tau^z_p$ and  $\tau^z_q$ are anti-aligned and the link $l$ is in the state $|\sigma\>$ or (2) $\tau^z_p$ and  $\tau^z_q$ are aligned, and the link $l$ is \emph{not} in the state $|\sigma\>$. In other words, $P_l$ projects onto states in which the $\sigma$ strings coincide with the \emph{domain walls} between the plaquette spins. 

\begin{figure}[b]
    \includegraphics[width=.45\textwidth]{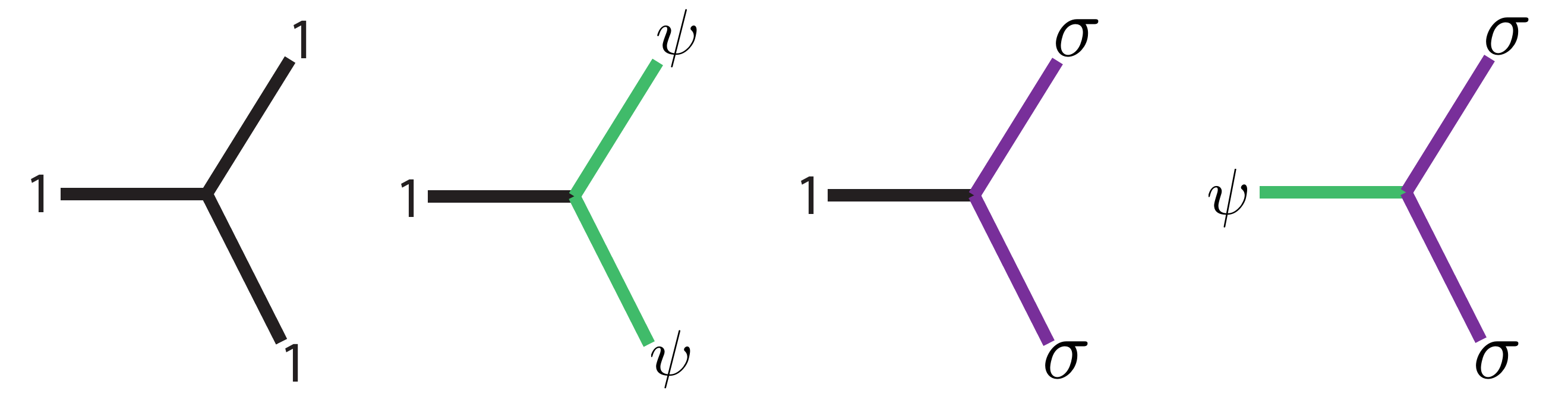}
        \centering
            \caption{The four types of vertices that obey the fusion rules. }
    \label{branching-rules}
    \end{figure}

The vertex operator $Q_v$ acts on the three spins living on the links adjacent to $v$  (blue spins in Fig. \ref{plaquette-operator}). It is defined by
\beq
Q_v    \left|
  \begin{minipage}[h]{0.09\linewidth}
        \vspace{0pt}
        \includegraphics[scale=.5]{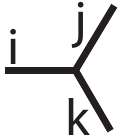}
   \end{minipage}\right\rangle=
 \delta_{ijk} \left|
  \begin{minipage}[h]{0.09\linewidth}
        \vspace{0pt}
        \includegraphics[scale=.5]{branching-delta.pdf}
   \end{minipage}\right\rangle\ 
\label{vertex-operator}
\eeq
where $\delta_{ijk}$ is a fully symmetric three-index tensor whose only nonzero elements are 
\begin{equation}
\delta_{111} = \delta_{1\psi\psi} = \delta_{1\sigma\sigma} = \delta_{\psi\sigma\sigma} = 1,
\end{equation}
together with cyclic permutations. In the string language, we will refer to the states with $\delta_{ijk} = 1$ as obeying the  `fusion rules', \ {and} those with $\delta_{ijk} = 0$ as violating the fusion rules (see Fig. \ref{branching-rules}). Thus, $Q_v$ is a projection operator that projects onto states that obey the fusion rules at $v$.

The $B_p$ operator acts on the plaquettes of the honeycomb lattice (red spins in Fig. \ref{plaquette-operator}) and is a linear combination of three terms: 
 \beq\label{bp}
B_p= a_1 B_p^1 + a_\psi B_p^\psi + a_\sigma B_p^\sigma \tau_p^x,
\eeq
where $a_1 = a_\psi = \frac{1}{4}$ and $a_\sigma = \frac{\sqrt{2}}{4}$. Each term, $B_p^s$, takes the form 
\beq\label{bps}
B_p^s=\mathcal{P}_p \tilde{B}_p^s\mathcal{P}_p
\eeq
where 
\beq
\mathcal{P}_p=\prod_{v\in p}Q_v
\eeq
is a projection operator which projects onto states that obey the fusion rules at the $6$ vertices adjoining $p$. Before giving formal definitions of these operators, it is worth noting that the $B_p^s$ operators are identical to the $B_p^s$ operators in the doubled Ising string-net model, a connection we will explore further in section \ref{relation-to-doubled-ising}.
%Before giving their definition, it is perhaps worth mentioning that the plaquette operators $B_p^0$, $B_p^\psi$, $B_p^\sigma$ are identical to the
%plaquette operators for the doubled-Ising string-net model. More generally, the model as a whole is very closely related to the doubled-Ising string-net model, as we will explain in section xxx.

\begin{figure}[tb]
      \centering
    \includegraphics[width=.44\textwidth]{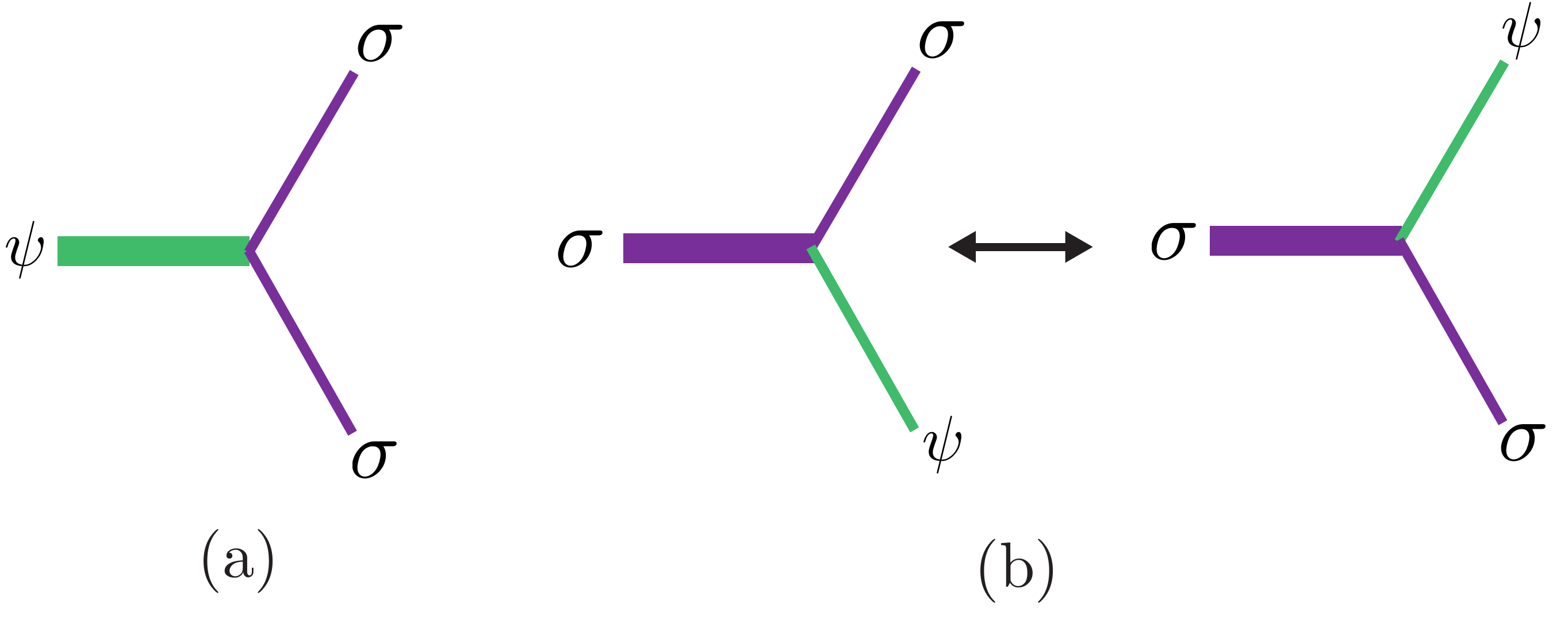}
            \caption{(a) The vertices counted by $N_{p1}$.
(b) The vertices counted by $N_{p2}$. For a vertex to be counted, its initial state must 
match the picture on the left and final state match the picture on the right or vice versa.
External legs of $p$ are shown in bold.}
    \label{Hvertices}
    \end{figure}

%All that remains is to define the three operators $\tilde{B}_p^0$, $\tilde{B}_p^\psi$ and $\tilde{B}_p^\sigma$. 
The first operator, $\tilde{B}_p^1$, is the simplest: 
\begin{equation}\label{bp1}
\tilde{B}_p^1 = \mathbf{1}. 
\end{equation}
The second operator, $\tilde{B}_p^{\psi}$, is more complicated. This operator acts on $12$ link spins --- $6$ of which lie along the boundary of the plaquette $p$ and $6$ of which lie on the `external legs' that adjoin $p$. Importantly, $\tilde{B}_p^\psi$ acts differently on the boundary spins than it does on the external leg spins: in particular, while $\tilde{B}_p^{\psi}$ changes the state of the boundary spins, it does not affect the state of the spins on the external legs. Thus, the nonvanishing matrix elements of $\tilde{B}_p^{\psi}$ can be parameterized as
\begin{equation}
   \left\langle
  \begin{minipage}[h]{0.15\linewidth}
        \vspace{0pt}
        \includegraphics[scale=.3]{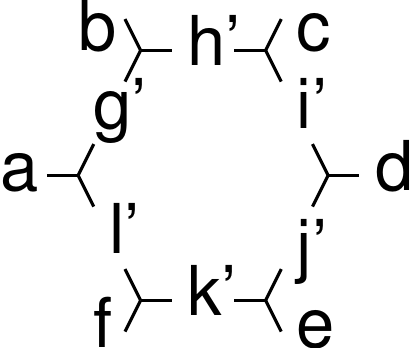}
   \end{minipage}\right |
     \tilde{B}_p^\psi
      \left|
  \begin{minipage}[h]{0.15\linewidth}
        \vspace{0pt}
        \includegraphics[scale=.3]{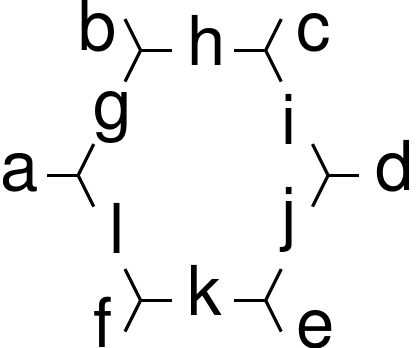}
   \end{minipage}\right\rangle = B_{p,ghijkl}^{\psi,g'h'i'j'k'l'}(abcdef)\non
\end{equation}
The explicit formula for these matrix elements is
\begin{align}\label{bppsi}
B_{p,ghijkl}^{\psi,g'h'i'j'k'l'}(abcdef) = \delta_{\psi gg'} \delta_{\psi hh'}\cdots \delta_{\psi ll'} \cdot (-1)^{N_{p1}}
\end{align}
where $N_{p1}$ denotes the number of vertices of $p$ of the type shown in Fig. \ref{Hvertices}a.
The matrix elements for $\tilde{B}_p^\sigma$ have a similar structure and are given by
\begin{align}\label{bpsigma}
B_{p,ghijkl}^{\sigma,g'h'i'j'k'l'}(abcdef) = \delta_{\sigma gg'} \cdots \delta_{\sigma ll'} \cdot 2^{-\frac{N_{p\sigma}}{4}}  (-1)^{N_{p2}} 
\end{align}
Here, $N_{p\sigma}$ denotes the number of external legs $a,b,\dots,f$ that are in the state $|\sigma\>$ while $N_{p2}$ denotes the number of vertices of $p$ for which the initial state has a vertex of the type shown on the left of Fig. \ref{Hvertices}b and a final state vertex of the type shown on the right of Fig. \ref{Hvertices}b or vice versa.

%----------------------------------------------
%----------------------------------------------
\subsection{Properties of the Hamiltonian}\label{hamiltonian-properties}

The Hamiltonian $H$ (\ref{toric-code-ham}) defined above has several interesting properties. Among the most important of these is that it possesses a global $\mz_2$ symmetry. This symmetry is defined by
\beq
S=\prod_p \tau^x_p,
\eeq
where $\tau^x_p$ denotes the Pauli operator acting on the plaquette spin $p$. We can think of $S$ as a conventional Ising symmetry which flips all the plaquette spins: $|+\> \leftrightarrow |-\>$. To see that $[S,H] = 0$, note that every term in $H$ either (1) does not act on the plaquette spins at all (i.e., $Q_v$, $B_p^1$, $B_p^\psi$), or (2) acts on the plaquette spins in a way that is manifestly symmetric under $S$ (i.e., $P_l$, $B_p^\sigma \tau_p^x$).

Another important property is that $H$ is a sum of mutually commuting operators. Indeed, from the expressions for $P_l$ and $Q_v$ given in Eq. \ref{link-projector} and \ref{vertex-operator} it is clear that
\beq
[P_l,P_{l'}]=[Q_v,Q_{v'}]=[P_l,Q_v]=0,
\eeq
since these operators are all diagonal in the $|\{\tau^z_p,\mu_l\}\>$ basis. It is also easy to see that
\beq
[B_p,Q_v]=[B_p,P_l]=0.
\eeq
The first equality follows from the fact that the plaquette operators include a factor of $\mathcal{P}$ on both sides. The second equality requires a little more work, but can be verified by separately considering each of the three terms in the definition of $B_p$. In particular, one can see that 
$B_p^1$ and $B_p^\psi$ both commute with $P_l$, since they commute with $(-1)^{N_{l\sigma}}$ for every link $l$. Likewise, one can see that $B_p^\sigma \tau^x_p$ commutes with $P_l$ by noting that $B_p^\sigma$ \emph{anti-commutes} with $(-1)^{N_{l\sigma}}$ if $l$ is adjacent to $p$ and commutes with $(-1)^{N_{l\sigma}}$ otherwise. The only relation left to establish is 
\begin{equation}
[B_p, B_{p'}] = 0
\end{equation}
The easiest way to prove this relation is to use the fact that the $B_p^s$ operators are identical to the string-net plaquette operators for the doubled Ising string-net model. String-net plaquette operators are known to commute with each other,\cite{string-net} so we know that $[B_p^s, B_{p'}^{s'}] = 0$, implying the above identity. 

A third property of  $H$ is that the operators $P_l, Q_v$ and $B_p$ are all \emph{projectors}. Indeed, we've already 
pointed this out for the case of $P_l$ and $Q_v$. As for $B_p$, the easiest way to see that it is a projector is to again invoke the fact that the $B_p^s$ 
operators are identical to string-net plaquette operators, which are known to obey the identity
\begin{equation}\label{projector-identity}
B_p^s B_p^{s'} = \sum_{t=1,\psi,\sigma} \delta_{ss't} B_p^t
\end{equation}
Using this identity, together with the fact that the coefficents $a_s$ obey the relation $\sum_{ss'} \delta_{ss't} a_s a_{s'} = a_t$, it is easy to check that $B_p^2 = B_p$, i.e. $B_p$ is a projector.  

Given that $H$ is a sum of commuting projectors, we know that the lowest energy states are those that obey
\beq
P_l|\Psi\>= Q_v|\Psi\> = B_p|\Psi\>=|\Psi\>, 
\label{gscond}
\eeq
for all links $l$, vertices $v$ and plaquettes $p$. In addition, we know that these states are separated from the excited states by a gap of at least $\Delta \geq 1$. Thus, the only property of the low energy spectrum left to determine is the ground state degeneracy of the model, i.e. the number of states that obey (\ref{gscond}). This degeneracy $D$ can be conveniently computed using the formula
\beq
D=\text{Tr}\left(\prod_l P_l \prod_v Q_v \prod_p B_p\right)
\eeq
We carry out this calculation in Appendix \ref{degeneracy}, and we find that the degeneracy depends on the global topology in which the model is defined. For example, if the model is defined in an infinite plane or spherical geometry, the ground state degeneracy is $D = 1$. In contrast, if the model is defined in a torus geometry, the ground state degeneracy is $D =  4$. More generally, we find that the ground state degeneracy on a surface of genus $g$ is $D = 4^g$. 

We can draw two conclusions from this calculation. First, we conclude that $H$ does not break the symmetry $S$ \emph{spontaneously} since the ground state is non-degenerate in an infinite plane or sphere geometry. Second, we conclude that $H$ realizes a topological phase (i.e. supports anyon excitations) since the ground state degeneracy is different for different topologies.

%----------------------------------------------
%----------------------------------------------
\subsection{Ground state wave function}

    \begin{figure}[tb]
    \includegraphics[width=.45
    \textwidth]{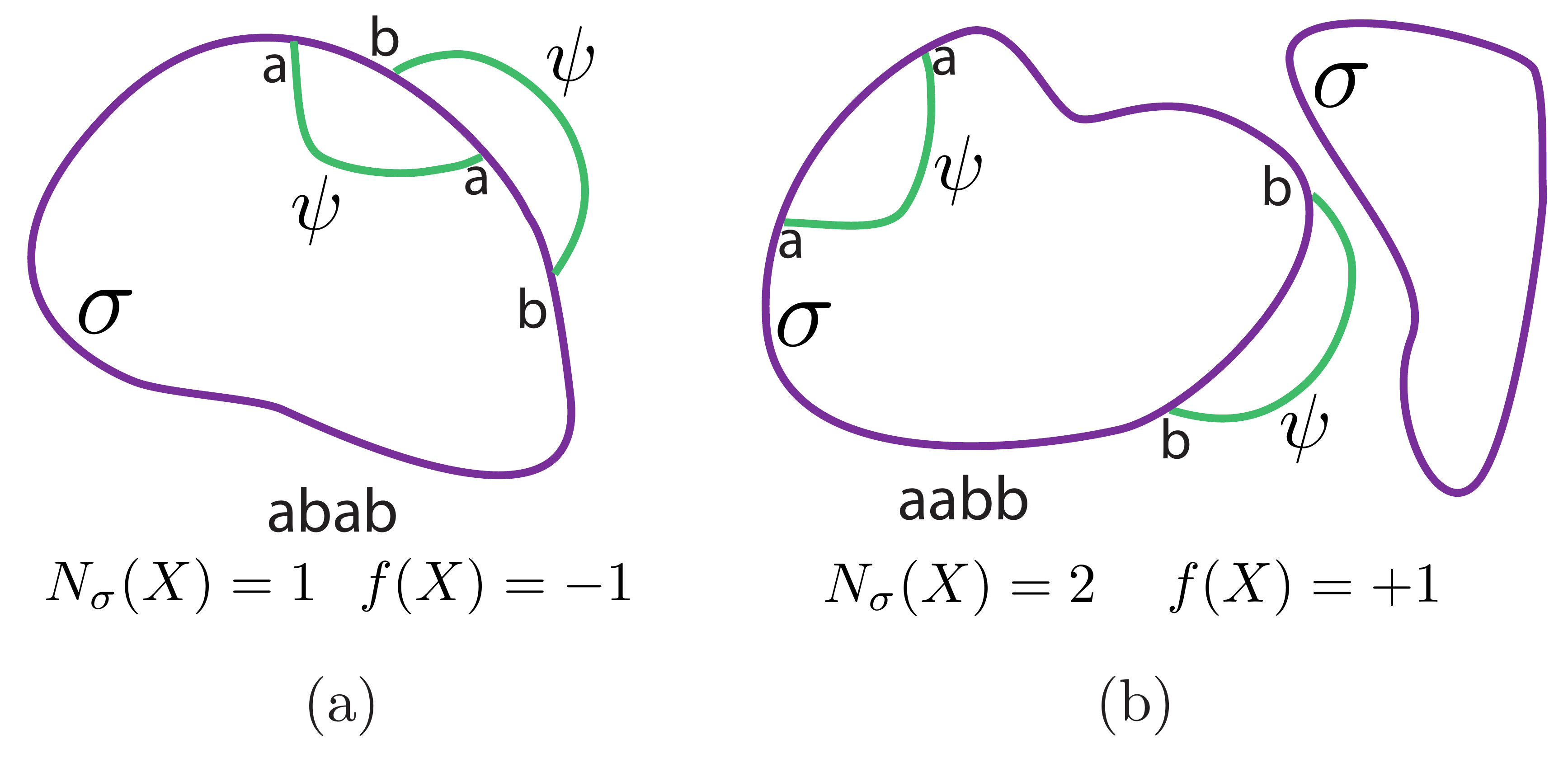}
        \centering
           \caption{(color online) Two examples of the function $f(X)$. Here the string states are drawn in the continuum rather than on the lattice. }
           \label{factor}
    \end{figure}
    
Interestingly, we can write down an explicit formula for the ground state wave function of the symmetric toric code model in an infinite plane or spherical geometry. Let $|X\> = |\{\tau^z_p, \mu_l\}\>$ be an arbitrary configuration of spins $\tau^z_p$ and strings $\mu_l$. The amplitude for $|X\>$ in the ground state vanishes unless $X$ satisfies two conditions: (1) the $\sigma$ strings lie along the domain walls of the $\tau^z_p$ spins, and (2) the $1, \psi, \sigma$ strings obey the fusion rules at each vertex (Fig. \ref{branching-rules}). Note that condition (2) implies that $\sigma$ strings form closed loops and $\psi$ strings either form closed loops or else their ends lie on $\sigma$ loops. If these conditions are satisfied, then the amplitude for $|X\>$ is given by\footnote{For simplicity, we do not bother to normalize the wave function.} 
\beq
\Psi(X) \equiv \<X |\Psi\> = \sqrt{2}^{N_{\sigma}(X)} f(X)
\label{gsform}
\eeq
where $N_{\sigma}(X)$ is the number of $\sigma$ loops contained in $X$ (Fig. \ref{factor}) and $f(X)$ takes the values $0$ or $\pm 1$. More specifically, $f(X)=0$ whenever there is at least one $\sigma$ loop in $X$ which has an odd number of $\psi$ strings ending on it, and $f(X) = \pm 1$ otherwise. Determining whether $f(X)=+1$ or $-1$ is a little bit trickier. To determine this sign, we compute a $\pm 1$ factor for each $\sigma$ loop in $X$ and then multiply them all together. The $\pm$ sign corresponding to each $\sigma$ loop can be calculated as follows. Suppose that there are $2n$ vertices where the $\psi$ strings end on the $\sigma$ loop. We can divide these vertices into two groups corresponding to the cases where the $\psi$ strings are incident from the outside of the loop or the inside of the loop. We then label the vertices in one group by `$a$', and the vertices in the other group by `$b$' (which one is which isn't important). We then go around the loop (in either direction) starting at some arbitrary point, reading off the sequence of $a$'s and $b$'s. We then count how many pairwise exchanges of $a$'s and $b$'s are necessary to rearrange the sequence so that $a$'s and $b$'s are separated into two blocks, i.e. $aa \cdots abb \cdots b$. The $\pm$ sign associated with the loop is given by the parity of the number of these exchanges (Fig. \ref{factor}). We explain how to derive Eq. (\ref{gsform}) in appendix \ref{gs-derivation}.

\iffalse
\begin{figure}[tb]%
    \centering
    \subfloat[]{{\includegraphics[width=.175\textwidth]{GS-factor1.pdf} }}%
    \qquad
    \subfloat[]{{\includegraphics[width=.245\textwidth]{GS-factor2.pdf} }}%
    \caption{ (color online) Two examples of string configurations with different $N_{\sigma}(X)$ and $f(X)$ . Here the string states are drawn in the continuum rather than on the lattice.}
    \label{factor}%
\end{figure}
\fi

%----------------------------------------------
%----------------------------------------------
\subsection{Excitations and string operators}\label{string-operators}

In this section, we show that the model supports four topologically distinct types of anyon excitations. We label these excitations by $\{1,\ e,\ m,\ em\}$, where
$1$ denotes the trivial excitation, and $e, m, em$ denote the three nontrivial excitations. We argue that this is a complete list and that there are no other topologically distinct anyons.

\begin{figure}[tb]%
    \centering
 \includegraphics[width=.4 \textwidth]{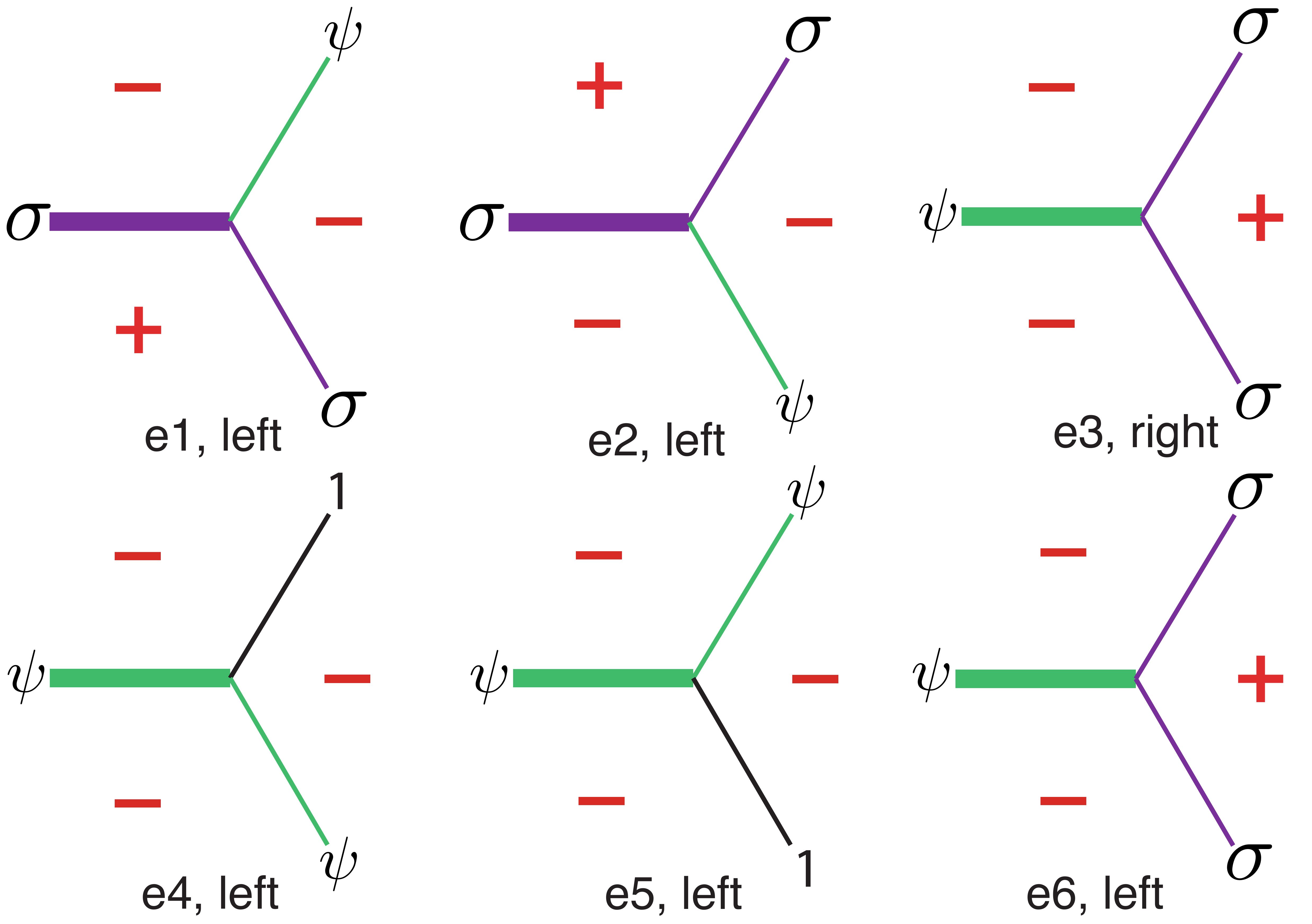}
    \caption{The vertices counted by $N_{e1},...,N_{e6}$. Vertices marked with `left' are only counted if the external leg (shown in bold)
adjoins $\gamma$ from the left, and similarly for those marked with `right.' The vertices counted by $N_{m1},...,N_{m6}$ can be obtained from the above set by flipping all the plaquette spins: $|+\>\leftrightarrow |-\>$.}
    \label{vertices}%
\end{figure}

We begin by describing the \emph{string} operators that create each of these anyon excitations (for a derivation of these operators see appendix \ref{string-operators-ungauged}). In general, these string operators act nontrivially along an open path $\gamma$, and when we apply them to the ground state, they create a pair of anyon excitations, with one at each end of the path $\gamma$.\cite{kitaev2003fault} The string operator that creates $e$-type anyons is defined by 
\begin{align}\label{w_e}
W_{e}^\gamma= 
\mathcal{P}_\gamma \cdot 
\prod_{l\in \gamma}(f_l)^{\frac{1}{4}(1+\tau_{p_l}^z)(1+\tau_{q_l}^z)} (-1)^{\sum_{i=1}^6 N_{ei}} 
\cdot \mathcal{P}_\gamma
\end{align}
where $\gamma$ is an (open) path on the honeycomb lattice and  $\mathcal{P}_\gamma$ denotes the projection operator
\begin{equation}\label{string-projector}
\mathcal{P}_\gamma = \prod_{v \in \gamma} Q_v \cdot \prod_{l \in \gamma} P_l
\end{equation}
Let us explain the notation in the formula for $W_e^\gamma$: the index $l$ runs over links that are contained in $\gamma$, while $p_l$ and $q_l$ denote the two plaquettes adjacent to link $l$. The link spin operator $f_l$ is defined as
\begin{equation}
f_{l} = |\psi\>\<1| + |1\>\<\psi| + |\sigma\>\<\sigma|
\end{equation}
 while the operators $N_{e1},...,N_{e6}$ count the number of vertices that belong to the path $\gamma$ and that are of the six types shown in Fig. \ref{vertices}. One subtlety is that the $N_{e1},...,N_{e6}$ operators distinguish between vertices where the external leg enters $\gamma$ on the `left' or on the `right', so to define these operators, we need to fix an orientation on $\gamma$. However the choice of orientation is not important: it is possible to show that changing the orientation of $\gamma$ only changes $W_{e}^\gamma$ by local operators acting near the ends of $\gamma$. 

Translating the formula for $W_e^\gamma$ into words, the action of $W^{\gamma}_e$ on a basis state $|\{\tau_p^z, \mu_l\}\>$ can be broken down into several steps. First, the projector $\mathcal{P}_\gamma$ annihilates the state unless (1) the strings obey the fusion rules for all the vertices $v$ in $\gamma$, and (2) the $\sigma$ strings lie along the domain walls of the $\tau^z$ spins, for all the links $l$ in $\gamma$. The next step is to multiply the state by a $\pm$ sign depending on whether $\gamma$ contains an even or odd number of vertices of the type shown in Fig. \ref{vertices}. The final step is to change the state of the link spins $\mu_l$ from $|1\>\rightarrow|\psi\>$  or $|\psi\rangle\rightarrow|1\rangle$ if the adjacent plaquette spins satisfy $\tau_{p_l}^z=\tau^z_{q_l}=+1$. 

The string operator for the $m$-type anyons takes a very similar form:
\begin{align}\label{w_m}
W_{m}^\gamma= \mathcal{P}_\gamma \cdot \prod_{l\in \gamma}(f_l)^{\frac{1}{4}(1-\tau_{p_l}^z)(1-\tau_{q_l}^z)} (-1)^{\sum_{i=1}^6 N_{mi}} \cdot \mathcal{P}_\gamma
\end{align}
This differs from the expression for $W_e^\gamma$ in two ways. First, $\tau_{p_l}^z, \tau_{q_l}^z$ are replaced by $-\tau_{p_l}^z, -\tau_{q_l}^z$, and second,
the $N_{ei}$ operators are replaced by $N_{mi}$. The $N_{mi}$ operators count the number of vertices that belong to $\gamma$ that have a particular type. These types are the same as those corresponding to $N_{ei}$ (Fig. \ref{vertices}), except with the plaquette spins flipped: $|+\> \leftrightarrow |-\>$. \footnote{Similar string operators were described by Ref. \onlinecite{schulz2015frustrated}, in the context of a phase with $\mz_2$ topological order and translation symmetry broken by densely packed $\sigma$ loops.}%A derivation of $W_e^{\gamma}$ and $W_m^{\gamma}$ is given in  appendix \ref{string-operators-ungauged}.

The only remaining string operator that we need to discuss is the one corresponding to $em$. As the notation suggests, this string operator can be obtained by multiplying together $W_e$ and $W_m$:
\begin{equation}
W_{em}^\gamma = W_e^\gamma W_m^\gamma
\end{equation}

To justify the above formulas, several points need to be established. First, we need to show that the above string operators only create excitations at their endpoints and nowhere else. Second, we need to show that the anyon excitations created by the string operators are topologically distinct. Finally, we need to show that the string operators create \emph{all} the topologically \ {distinct} excitations. 

To establish the first point we note that the above string operators have the property that they commute with every term in the Hamiltonian except for the six $B_p$ terms that act on the plaquettes adjoining the endpoints of $\gamma$. This property can be established either by straightforward (but tedious) algebra, or by using a graphical representation of the string operators as in Ref. \onlinecite{string-net}. As for the second point, this follows from the braiding statistics calculation presented in the next section: in particular, our calculation shows that the $e, m$ and $em$ excitations all have distinct braiding statistics and are therefore topologically distinct. Finally, to establish the last point, we recall that the ground state degeneracy of the model in a torus geometry is $4$. Typically, the ground state degeneracy on a torus is equal to the number of distinct anyon types\footnote{This relation can break down for some systems with non-Abelian anyons, or systems where translational symmetry permutes anyon excitations. We assume that our model does not fall into one of these categories.} so we conclude that $\{1,e,m,em\}$ is a complete list of anyons.

\ {A final remark: it is interesting to note that the above string operators have especially simple behavior when acting on states with no $\sigma$ strings and with $\tau_p^z = +1$ everywhere. Indeed, if we denote these states by $|+,\mu_l\rangle$, we can see from Eq. (\ref{w_e}), that the action of $W_e^\gamma$ on $|+,\mu_l\rangle$ is simply to flip the link spins from $|1\> \leftrightarrow |\psi\>$ on all the links along $\gamma$. Likewise, the action of $W_m^{\gamma}$ on $|+,\mu_l\rangle$ is to multiply the state by a factor of $(-1)^{N_{\psi}}$, where $N_{\psi}$ counts the number of $\psi$ strings adjoining $\gamma$ on the left.} 
%Putting these together, we can see that the action of $W_{em}^{\gamma}$ on $|+,\mu_l\rangle$  flips the link spins $1\leftrightarrow \psi$ along $\gamma$ and multiplies the state by $(-1)^{N_{\psi}}$. }

\ {What is particularly interesting is that these formulas precisely agree with the action of the $W_e$, $W_m$ string operators in the usual toric code string-net model built from string types $\{1,\psi\}$. Furthermore, if we consider symmetry reversed states, $|-,\mu_l\rangle$, where $\tau_p^z = -1$ for all $p$, this correspondence is reversed. That is, the action of $W_e^\gamma$ on $|-, \mu_l\>$ agrees with the action of $W_m$ in the usual toric code and vice versa. This connection between the string operators in the symmetric toric code and usual toric code is not a coincidence and holds more generally for all the symmetry enriched models that we describe below. }

%----------------------------------------------
%----------------------------------------------

\subsection{Symmetry action on anyons}
With expressions for the $e$ and $m$ string operators in hand, we can derive one of the most interesting features of our model: the $\mz_2$ symmetry $S$ \emph{exchanges} $e$-type anyons and $m$-type anyons. To see this, note that the string operators $W_e^\gamma, W_m^\gamma$ (\ref{w_e}-\ref{w_m}) are related to one another by
\beq \label{symmetry-action}
S^{-1} W_e^\gamma S=W_m^{\ {\gamma}}, \quad S^{-1} W_m^\gamma S = W_e^{\ {\gamma}}
\eeq
We conclude that $S$ \emph{exchanges} $e$-type anyons and $m$-type anyons. On the other hand, $S^{-1} W_{em}^{\ {\gamma}} S = W_{em}^{\ {\gamma}}$, so the symmetry leaves $em$-type particles unchanged. 

We note that a toric code with a symmetry exchanging $e$ and $m$ has been studied previously. \cite{wen2003quantum,Teo_Hughes} However, in that case, the symmetry was not an onsite $\mz_2$ symmetry, but rather a spatial symmetry.

%----------------------------------------------
%----------------------------------------------

\subsection{Braiding statistics}

\begin{figure}[tb]
    \includegraphics[width=.4\textwidth]{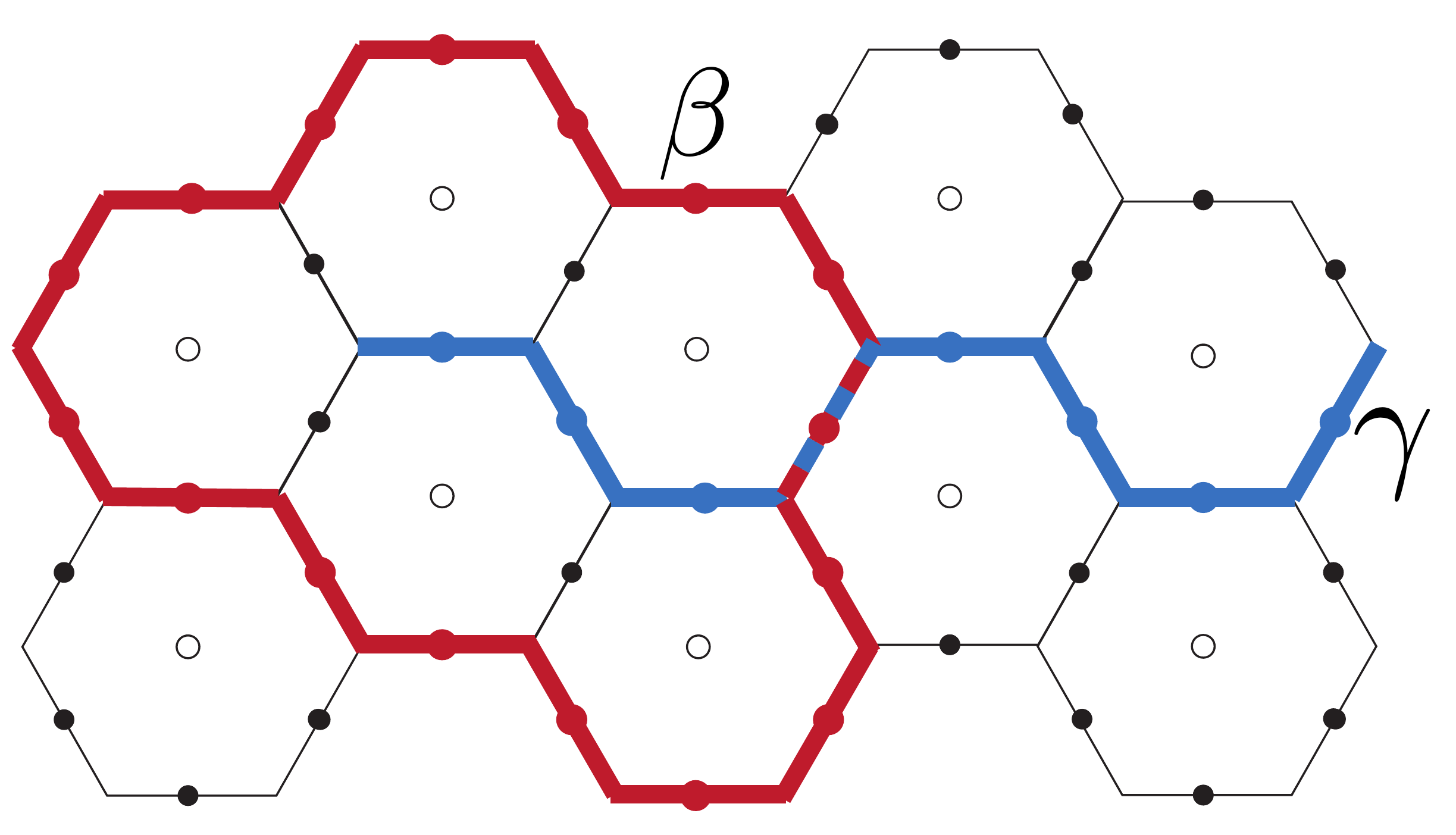}
        \centering
          \caption{An example of paths $\beta, \gamma$ used to find the mutual statistics of $e$ and $m$. Applying the two string operators 
in two different orders gives results that differ by the mutual statistical phase $e^{i \theta_{e,m}}$.}
             \label{braiding}
    \end{figure}

We now compute the braiding statistics data for the $\{1, e, m, em\}$ anyons, and we show that this data agrees with that of the toric code model. We begin with the mutual statistics of the anyons --- that is, the statistical phase $e^{i\theta_{a,b}}$ associated with braiding anyon $a$ around anyon $b$ where $a, b \in \{1, e, m, em\}$. We compute this statistical phase using a standard approach:\cite{kitaev2003fault, LevinWenHop, string-net} we construct the string operators associated with $a$ and $b$, with one acting on the path $\beta$ and the other on the path $\gamma$ (Fig. \ref{braiding}), and then we derive the mutual statistics from the string operator commutation algebra using the general relation:
\beq
W_a^\beta W_b^\gamma =e^{i\theta_{a,b}}W_b^\gamma W_a^\beta
\label{comm-alg-stat}
\eeq
We start with the mutual statistics of the $e$ and $m$ type anyons. To find their statistics, we need to compare $W_m^\beta W_e^\gamma$ with $W_e^\gamma W_m^\beta$. To make a comparison, it is easiest to consider the action of these operators on a particular state, say $|\{\tau^z_p = +,\mu_l = 1\}\>$. First we apply $W_e^\gamma$. The result is to flip the link spins from $|1\>$ to $|\psi\>$ along the path $\gamma$:
\begin{equation*}
W_e^{\gamma}|\{+,1\}\> = |\{\tau^z_p = +,\mu_{l\in\gamma}=\psi,\mu_{l\notin \gamma}=1\}\>
\end{equation*}
Next we apply $W_m^\beta$. This has no effect on the link spins since $W_m^\beta$ only flips the link spins when $\tau^z_p = -$. On the other hand, applying $W_m^\beta$
has the effect of multiplying the state by $-1$ since $\prod_{l \in \beta} (-1)^{\sum_i N_{mi}} = -1$ for the above state. Thus, we have
\beq\label{me-comm}
W_m^{\beta} W_e^{\gamma}|\{+,1\}\> =-|\{\tau^z_p = +,\mu_{l\in\gamma}=\psi,\mu_{l\notin \gamma}=1\}\>
\eeq
Now consider the opposite ordering. First, we apply $W_m^\beta$ to $|\{+,1\}\>$. In this case, the state is left unchanged:
\begin{equation*}
 W_m^{\beta}|\{+,1\}\> = |\{\tau^z_p = +,\mu_l = 1\}\>
\end{equation*}
Note that there is no factor of $-1$ here because $\prod_{l \in \beta} (-1)^{\sum_i N_{mi}} = +1$ for the above state. If we now apply $W_e^\gamma$, we obtain
\beq\label{em-comm}
W_e^{\gamma} W_m^{\beta} |\{+,1\}\> = |\{\tau^z_p = +,\mu_{l\in\gamma}=\psi,\mu_{l\notin \gamma}=1\}\>
\eeq
Again, there is no factor of $-1$ since $\prod_{l \in \beta} (-1)^{\sum_i N_{ei}} = +1$.

Comparing Eqs. \ref{em-comm} and \ref{me-comm}, we can see that $W_e^\gamma$ and $W_m^\beta$ \emph{anti-commute} when acting on the state $ |\{+,1\}\>$. In fact, this anti-commutation property is more general: these operators anti-commute when acting on \emph{any} state. We have verified this by expressing the string operators in matrix form and computing their anti-commutator numerically. In view of Eq. (\ref{comm-alg-stat}), the implication of this result is that the $e$ and $m$ type particles have a mutual statistical phase of
\begin{equation}
e^{i \theta_{e,m}} = -1
\end{equation}
Following a similar analysis, one can show that $W_e^{\gamma}$ and $W_e^\beta$ \emph{commute} with one another (as do $W_m^{\gamma}$ and $W_m^\beta$) so that the corresponding statistical phases are
\begin{equation}
e^{i \theta_{e,e}} = e^{i \theta_{m,m}} = +1
\end{equation}
All of these phases agree with the toric code model.\cite{kitaev2003fault} Furthermore, the statistical phases involving the $em$ particle also agree since these are fully determined by the fact that $em = e \times m$ together with the above data. 

The only thing left to check is that the anyons in our model have the same \emph{exchange} statistics as those in the toric code model. In particular, we need to check that the $e$ and $m$ particles are bosons, while the $em$ particle is a fermion. Conveniently we can establish this fact without doing any additional calculation. To see this, note that $e$ and $m$ have to be either bosons or fermions, given the mutual statistics found above. Furthermore symmetry requires that $e$ and $m$ have the same exchange statistics. Thus, the only alternative possibility is that $e$ and $m$ are both fermions, in which case $em$ is also a fermion --- by the composition rule for exchange statistics. But we can rule out this three-fermion possibility by noting that if $e, m, em$ were all fermions, then the chiral central charge $c_{-}$ of our model would have to be $c_{-} \equiv 4 \pmod{8}$, \footnote{This follows from a general relation that holds for any topological phase, namely $e^{2\pi i c_-/8} = \frac{1}{D}\sum_s d_s^2 \theta_s$ where $D = \sqrt{\sum_s d_s^2}$. Here the sum runs over all anyons $s$. and $\theta_s, d_s$ denote the topological spin and quantum dimension of $s$. See e.g. Ref. \onlinecite{Kitaev2006} and references therein.} yet we know that the chiral central charge of our model must vanish since the Hamiltonian is a sum of commuting projectors. %To summarize, we have shown that ${1,e,m,em}$ have the same exchange statistics, mutual statistics, and fusion rules as the toric code.

%%%%%%%%%%%%%%%%%%%%%%%%%%%%%%%%
%%%%%%%%%%%%%%%%%%%%%%%%%%%%%%%%
%%%%%%%%%%%%%%%%%%%%%%%%%%%%%%%%
%%%%%%%%%%%%%%%%%%%%%%%%%%%%%%%%

\section{Relationship to doubled Ising string-net model}\label{relation-to-doubled-ising}

In this section we discuss the relationship between the symmetric toric code and the doubled Ising string-net model. We present two results. First, we show that if we gauge the global $\mz_2$ symmetry of the symmetric toric code model, the resulting gauge theory can be mapped \emph{exactly} onto a variant of the doubled Ising string-net model. Second, we reverse the logic and we show that the symmetric toric code model can be constructed from the doubled Ising string-net model by applying an `ungauging' procedure.

Why are these results important? As we explain below, our first result provides an alternative proof that the symmetric toric code model has the properties claimed above, namely (1) it supports anyon excitations with the same braiding statistics as in the conventional toric code, and (2) the $\mz_2$ symmetry exchanges the $e$ and $m$ type particles. Likewise, our second result is significant because it shows how the symmetric toric code model was constructed. Most important of all, both of these results generalize to arbitrary symmetry enriched string-net models and they lie at the heart of our construction and analysis of the more general models in section \ref{general-framework}.

%Indeed, it is known\cite{stationq} that properties (1) and (2) are guaranteed to hold for any $\mz_2$ symmetric system that, upon gauging the symmetry, gives a model belonging to the doubled Ising phase. 

\subsection{Review of doubled Ising string-net model}\label{doubled-Ising}

We begin with a brief review of the doubled Ising string-net model. (See Ref. \onlinecite{string-net} for a general introduction to the string-net formalism). The doubled Ising string-net model is a spin system built out of spins that live on the links of the honeycomb lattice. Each spin can be in \emph{three} states, which we denote by $|1\>$, $|\psi\>$, and $|\sigma\>$. The states of the spins can be equivalently described in terms of `strings': if a spin is in the state $|\psi\>$ or $|\sigma\>$, we will say that the corresponding link is occupied by a $\psi$ or $\sigma$ string, while if the spin is in the state $|1\>$, we will say the corresponding link is unoccupied. The Hamiltonian of the model is given by
\beq\label{doubled-ising-ham}
H^{di}=-\sum_v Q_v -\sum_p B_p^{di}
\eeq
The first term, $Q_v$, acts on three links adjacent to a vertex $v$, and is defined as in Eq. (\ref{vertex-operator}). This term energetically favors vertices that satisfy the fusion rules shown in Fig. \ref{branching-rules}. The second term, $B_p^{di}$ can be written as a sum
\beq
B_p^{di} = a_1 B_p^1+ a_\psi B_p^{\psi}+ a_\sigma B_p^{\sigma}
\eeq
where $a_1 = a_\psi = \frac{1}{4}$ and $a_\sigma = \frac{\sqrt{2}}{4}$. The operators $B_p^1, B_p^\psi, B_p^\sigma$ each act on 12 links --- 6 of which make up the boundary of a plaquette $p$ and another 6 that lie on the external legs that adjoin $p$. These operators are defined as in Eqs.  (\ref{bps}-\ref{bpsigma}). (For readers who are familiar with string-net models, we should mention that the definitions of $B_p^1, B_p^\psi, B_p^\sigma$ given in Eqs.  (\ref{bps}-\ref{bpsigma}) are equivalent to the standard definitions of $B_p^s$ in terms of $F$-symbols).

The ground state of the model, $|\Psi^{di}\>$, is a linear superposition of string configurations that obey the fusion rules. The amplitude for each configuration can be computed using the following `local rules':
\begin{align}
 \Psi^{di}
\bpm \includegraphics[height=0.3in]{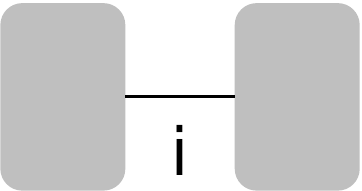} \epm  =&
\Psi^{di} 
\bpm \includegraphics[height=0.3in]{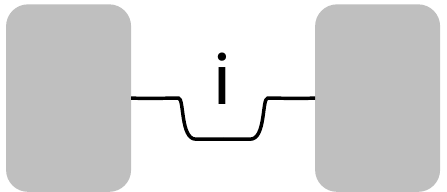} \epm
\label{topinv}
\\
 \Psi^{di}
\bpm \includegraphics[height=0.3in]{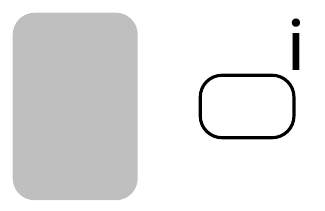} \epm  =&
d_i\Psi^{di} 
\bpm \includegraphics[height=0.3in]{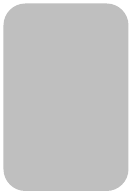} \epm
\label{clsdst}
\\
 \Psi^{di}
\bpm \includegraphics[height=0.3in]{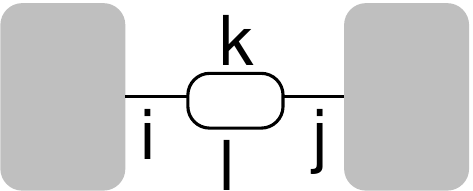} \epm  =&
\delta_{ij}
\Psi^{di} 
\bpm \includegraphics[height=0.3in]{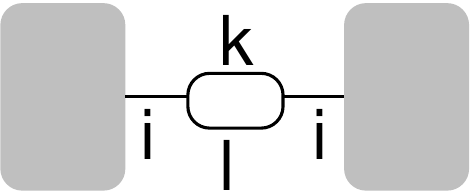} \epm
\label{bubble}\\
 \Psi^{di}
\bpm \includegraphics[height=0.3in]{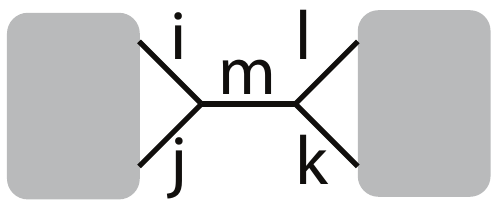} \epm  =&
\sum_{n} 
F^{ijm}_{kln}
\Psi^{di} 
\bpm \includegraphics[height=0.28in]{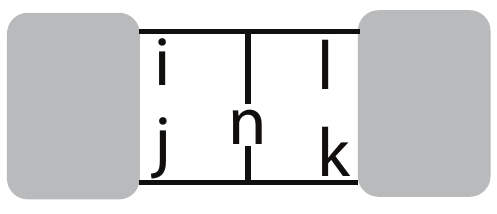} \epm
\label{fusion} 
\end{align}
Here $d_i$ is the quantum dimension of the $i$'th edge label
\beq
d_1=d_\psi=1,\ d_\sigma=\sqrt{2}
\eeq
while $F^{ijm}_{kln}$ is the `$F$-symbol' for the doubled Ising model, with the following nonzero components:
\begin{align}\label{F-symbols}
&F^{211}_{211}=F^{121}_{121}=-1, \ F^{11j}_{11k}=\frac{(-1)^{\frac{j\cdot k}{4}}}{\sqrt{2}}, \ j,k=0,2\non \\
&F^{220}_{220}=F^{220}_{111}=F^{110}_{221}=F^{121}_{210}=F^{211}_{120}=F^{121}_{101}=F^{211}_{011}=1\non \\
&F^{121}_{012}=F^{211}_{102}=F^{011}_{122}=F^{101}_{212}=F^{101}_{121}=F^{011}_{211}=F^{000}_{000}=1\non \\
&F^{i0i}_{i0i}=F^{0ii}_{0ii}=F^{000}_{iii}=F^{ii0}_{00i}=F^{i0i}_{0i0}=F^{0ii}_{i00}=1, \ \ i=1,2
\end{align}
where
\beq\label{number-labels}
0=\text{vacuum}, \ \ 1=\sigma, \ \ 2=\psi
\eeq

Let us specialize to the case where the model is defined in a sphere or infinite plane geometry. In this case, the above local constraint equations uniquely determine the ground state wave function $\Psi^{di}$ since they allow us to relate the amplitude of any string configuration to the amplitude of the vacuum (empty string) \ {configuration} which is defined to be $\Psi^{di}(\text{vacuum}) = 1$, by convention. Using this approach, it is not hard to show that the ground state amplitude is given by $\Psi^{di}(X) = \sqrt{2}^{N_{\sigma}(X)} f(X)$, in exact agreement with Eq.  (\ref{gsform}). This agreement is not a coincidence: in fact, the way we derive Eq. (\ref{gsform}) is to first find the ground state wave function of the doubled Ising string-net model and then use the connection with the symmetric toric code to obtain the ground state of the symmetric toric code model. See appendix \ref{gs-derivation} for details.

Physically, the most important property of the doubled Ising string-net model is that it supports $9$ different types of anyon excitations. These anyons can be labeled as:
\beq
\{1,\psi,\sigma\}\times\{1,\bar{\psi},\bar{\sigma}\}=\{1,\psi,\bar{\psi},\sigma,\bar{\sigma},\psi\bar{\psi},\sigma\bar{\psi},\psi\bar{\sigma},\sigma\bar{\sigma}\}\non
\eeq
The braiding statistics and fusion rules of these anyons are described by the $\text{Ising} \times\overline{\text{Ising}}$ topological phase. This is why the model is called the `doubled Ising' string-net model. 

%is as well as the string operators that create them, can be obtained using the string-net formalism. For our purposes, we will only need to consider the string operators corresponding to the $\sigma\bar{\sigma}$ and $\psi \bar{\psi}$ particles which we will discuss in appendix \ref{string-operators-ungauged}.

    \begin{figure}[tb]
    \includegraphics[width=.25\textwidth]{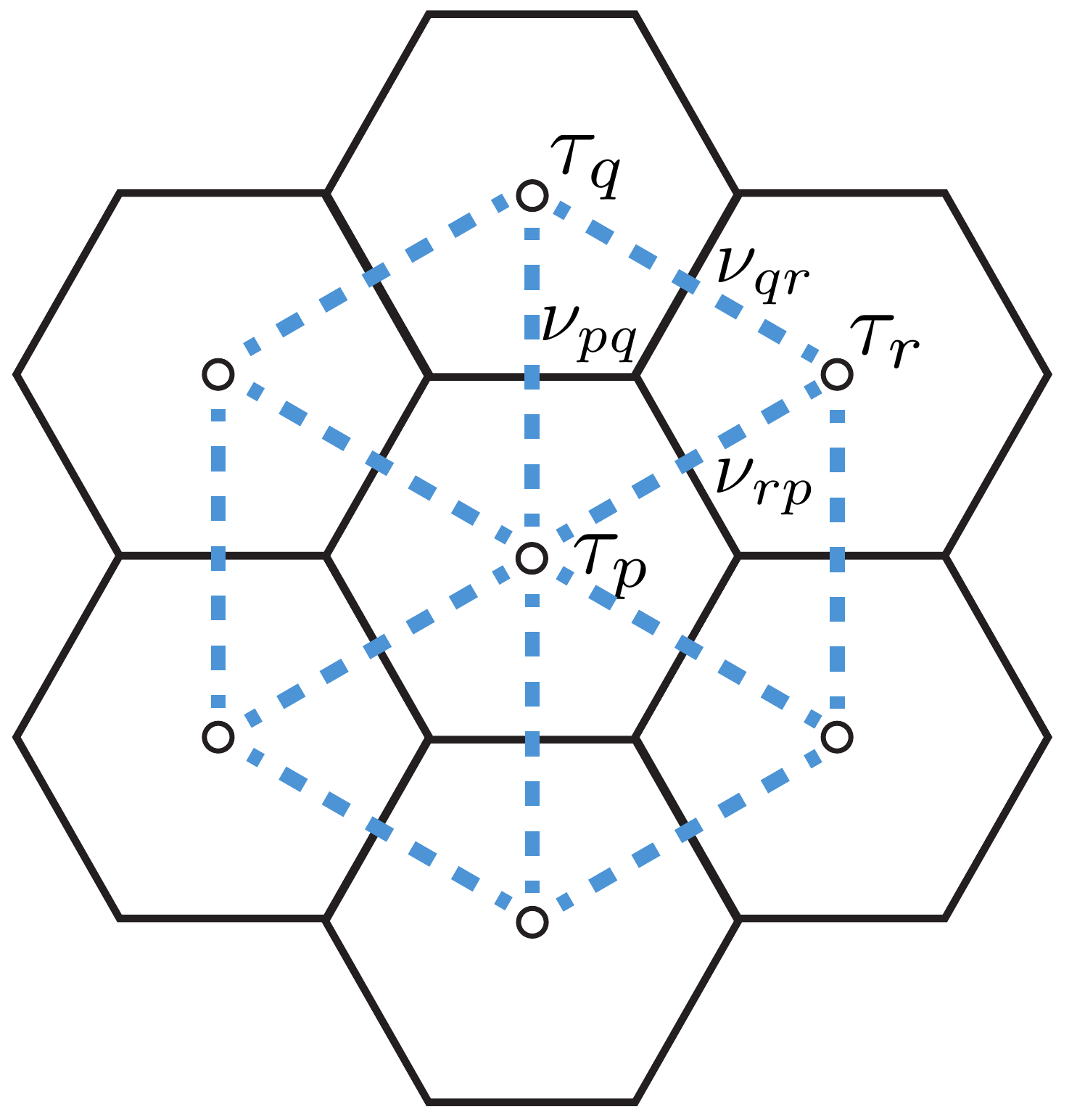}
        \centering
          \caption{The gauge field degrees of freedom, $\nu_{pq}$, live on the links $\<pq\>$ of the dual \emph{triangular} lattice connecting neighboring plaquette spins (we omit link spins for clarity).}
    \label{gauge-dof}
    \end{figure}

%\subsection{Gauging the symmetry}\label{gauging}

%In this section we investigate what happens if we gauge the global $\mz_2$ symmetry of our toric code model. Our main result is that the gauged toric code model belongs to the same phase as the doubled Ising string-net model. 

%We establish this by finding a local unitary transformation that maps the ground state of the gauged model onto the ground state of the doubled Ising string-net model, tensored with a product state.

\subsection{The gauged toric code model}\label{gauging}

Setting aside the doubled Ising string-net model for the moment, we now explain how to gauge the global $\mz_2$ symmetry of the symmetric toric code model. Following the standard procedure,\cite{kogut79} the first step is to add a gauge field to the model. This gauge field should be placed on links connecting neighboring lattice sites where the symmetry acts. In our case, the symmetry $S = \prod_p \tau_p^x$ only acts on the plaquette spins, so we put the gauge field on the links $\<pq\>$ of the \emph{triangular} lattice connecting the plaquette spins (Fig. \ref{gauge-dof}). Since the gauge group is $\mz_2$, the lattice gauge field is a two-state degree of freedom $\nu_{pq}^z = \pm 1$. 
    
The second step is to perform the minimal coupling procedure, replacing $\tau_p^z\tau_q^z\rightarrow\tau_p^z\nu_{pq}^z\tau_q^z$, whenever there are terms in the Hamiltonian that couple neighboring plaquette spins $\tau_p^z, \tau_q^z$. The final step is to add to the Hamiltonian the term $-\sum_{\langle pqr\rangle}\nu_{pq}^z\nu_{qr}^z\nu_{rp}^z$, which ensures that the states with zero $\mz_2$ gauge flux have the lowest energy. Applying these steps to the symmetric toric code model gives the following gauged Hamiltonian:
\begin{equation}
H^{\text{gauged}} = -\sum_v Q_v-\sum_p B_p - \sum_l P^{\text{gauged}}_l - \sum_{\langle pqr\rangle}\nu_{pq}^z\nu_{qr}^z\nu_{rp}^z
\end{equation}
where
\begin{equation}
P^{\text{gauged}}_l = \frac{1}{2}(1+(-1)^{N_{l\sigma}} \cdot \tau^z_p\nu_{pq}^z \tau^z_q)
\end{equation}
Notice that $Q_v$ and $B_p$ are not affected by the minimal coupling procedure since they don't contain any terms like $\tau_p^z \tau_q^z$.
Like any gauge theory, the Hamiltonian $H^{\text{gauged}}$ is defined on a Hilbert space consisting of \emph{gauge invariant} states, that is, states $|\psi\>$ obeying 
\beq\label{gauge-invariance}
\tau_p^x \prod_q\nu_{pq}^x |\psi\> = |\psi\>
\eeq
for all $p$.

As an aside, we should mention that we skipped one of the steps in the usual gauging procedure. This step involves multiplying each term in the gauged Hamiltonian %(e.g. $P_l, Q_v, B_p$)
by an operator that projects onto states that have vanishing $\mz_2$ gauge flux through all triangular plaquettes contained in the region of support of the term in question.\cite{levin2012braiding,wangcj15}
%This step has two purposes: (1) it guarantees that the gauged Hamiltonian is Hermitian and (2) it makes the gauging procedure unambiguous even in the case where the Hamiltonian contains further neighbor couplings $\tau_p^z \tau_q^z$.
While this step is important when gauging a general spin model, it is not necessary when the Hamiltonian only includes nearest neighbor $\tau_p^z \tau_q^z$ couplings, as is the case here.
%------------------------------------

%------------------------------------
\subsection{Equivalence between gauged toric code and doubled Ising string-net model}\label{mod-string-net}

In this section, we establish an exact equivalence between the gauged toric code model and a \emph{modified} version of the doubled Ising string-net model. Using this equivalence, we then present an alternative proof that the symmetric toric code model has the topological and symmetry properties claimed above. 

To begin, we explain how the modified doubled Ising string-net model is defined. The modified model differs from the standard Ising string-net model $H^{di}$ in two ways. First, the Hilbert space is bigger: in addition to the three-state spin $\mu_l = 1, \psi, \sigma$, the model also has a two-state spin $\xi_l^z= \pm$ living on each link $l$ of the honeycomb lattice. Also, the Hamiltonian has two extra terms:
\begin{align}
H^{di'} = H^{di} - \sum_l \frac{1}{2}(1+\xi_l^z) - \sum_v \prod_{l \in v} \xi_l^z (-1)^{N_{l \sigma}}
\label{modising}
\end{align}
Despite these differences, the ground states of the two models, $\Psi^{di'}$, $\Psi^{di}$, are almost the same:
\begin{equation}
|\Psi^{di'}\> = |\Psi^{di}\> \otimes |\{\xi_l^z = +\}\>
\label{gsrel}
\end{equation}
To see this, notice that $|\Psi^{di}\> \otimes |\{\xi^z = +\}\>$ separately minimizes the energy of every term in $H^{di'}$. 

The fact that the two ground states $|\Psi^{di'}\>$, $|\Psi^{di}\>$ are identical up to tensoring with a product state is very important for us because it implies that the two models have the same topological properties, i.e. the same anyon excitations and the same braiding statistics. The two models are thus interchangeable for our purposes.

We now show that the modified string-net model $H^{di'}$ is exactly equivalent to the gauged toric code model. To this end, we need to construct a unitary mapping $U$ between the Hilbert spaces of the two models that transforms the Hamiltonian $H^{di'}$ into the Hamiltonian $H^{\text{gauged}}$. The easiest way to define the mapping $U$ is in terms of basis states. The Hilbert space of the modified string-net model is spanned by basis states of the form $|\{\xi_l^z, \mu_l\}\>$ where $\xi_l^z = \pm$ and $\mu_l = 1, \psi, \sigma$. Likewise, the Hilbert space of the gauged model is spanned by basis states $|\{\tau_p^z \nu_{pq}^z \tau_q^z, \mu_l\}\>$ labeled by the gauge invariant quantum numbers $\tau_p^z \nu_{pq}^z \tau_q^z = \pm$ and $\mu_l = 1, \psi, \sigma$. With this notation, the mapping $U$ is defined by 
\begin{align}
U |\{\tau_p^z \nu_{pq}^z \tau_q^z, \mu_l\}\> = |\{\xi_l^z, \mu_l\}\>
\label{mapping}
\end{align}
where $\xi_l^z = (-1)^{N_{l\sigma}} \tau_p^z \nu_{pq}^z \tau_q^z $ and $l$ is the link separating the two plaquettes $p,q$ (see Table \ref{mapping-table}). To see that $U$ transforms $H^{di'}$ into $H^{\text{gauged}}$, note that
\begin{align}
&U^{-1} \xi_l^z U = (-1)^{N_{l\sigma}}\tau_p^z \nu_{pq}^z \tau_q^z , \quad U^{-1} Q_v U = Q_v, \nonumber \\ 
&U^{-1} N_{l \sigma} U = N_{l \sigma}, \quad U^{-1} B_p^1 U = B_p^1, \nonumber \\
&U^{-1} B_p^\psi U = B_p^\psi, \quad U^{-1} B_p^\sigma U = B_p^\sigma \tau_p^x
\end{align}
Substituting these formulas into (\ref{modising}) gives 
\beq
U^{-1} H^{di'} U = H^{\text{gauged}},
\eeq
 as claimed.

The equivalence between the gauged toric code model and the modified doubled Ising string-net model provides another proof that the (ungauged) toric code model has the properties claimed above, i.e. (1) it has the same anyon and braiding statistics as the conventional toric code, and (2) the $\mz_2$ symmetry exchanges the $e$ and $m$ particles. The reason this is so is that the mapping between ungauged and gauged models is known to be one-to-one in the sense that the braiding statistics data for the excitations of the gauged model \emph{uniquely} determines the braiding statistics and symmetry data for the ungauged model.\cite{stationq,etingof} In particular, if the excitations of a gauged $\mz_2$ symmetric model are described by the $\text{Ising} \times \overline{\text{Ising}}$ topological phase, as is the case here, then it is known that the ungauged model must obey properties (1) and (2).\cite{gu2014unified, stationq}

%Indeed, it is known\cite{stationq} that properties (1) and (2) are guaranteed to hold for any $\mz_2$ symmetric system that, upon gauging the symmetry, gives a model belonging to the doubled Ising phase. 

\begin{table}
\begin{tabular}{ l  | r }
  \hline
Gauged toric code & Modified String-net  \\  \hline
  \renewcommand{\arraystretch}{1.5}
   $|\tau_p^z\nu_{pq}^z\tau_q^z =+1,\ \mu_l=1\>$ &\   $|+,1\>$ \\
 \  $|\tau_p^z\nu_{pq}^z\tau_q^z =+1,\ \mu_l=\psi\>$ &\  $|+,\psi\>$    \\
   \  $|\tau_p^z\nu_{pq}^z\tau_q^z =-1,\ \mu_l=\sigma\>$ &\ $|+,\sigma\>$   \\
   \hline 
 \  $|\tau_p^z\nu_{pq}^z\tau_q^z =-1,\ \mu_l=1\>$  &\  $|-,1\>$   \\
  \  $|\tau_p^z\nu_{pq}^z\tau_q^z =-1,\ \mu_l=\psi\>$ &\ $|-,\psi\>$     \\
   \  $|\tau_p^z\nu_{pq}^z\tau_q^z =+1,\ \mu_l=\sigma\>$ &\  $|-,\sigma\>$  \\
 \hline
\end{tabular}
\caption{The mapping $U$ (\ref{mapping}) between the Hilbert space of the gauged toric code model and the Hilbert space of the modified string-net model.}
\label{mapping-table}
\end{table}

%% -- commented out below is a previous version of this table using different notation
\iffalse
\begin{table}
\begin{tabular}{ l  | r }
  \hline
String-net & Gauged toric code \\  \hline
  \renewcommand{\arraystretch}{1.5}
  $|1\>$ &\ $\mu_l=1$,  $\tau_p^z\nu_{pq}^z\tau_q^z =+1$\\
 \ $|\psi\>$ &\ $\mu_l=\psi$,  $\tau_p^z\nu_{pq}^z\tau_q^z =+1$  \\
   \ $|\sigma\>$ &\ $\mu_l=\sigma$,  $\tau_p^z\nu_{pq}^z\tau_q^z =-1$ \\
 \ $|x\>$ &\ $\mu_l=1$,  $\tau_p^z\nu_{pq}^z\tau_q^z =-1$   \\
  \ $|y\>$ &\ $\mu_l=\psi$,  $\tau_p^z\nu_{pq}^z\tau_q^z =-1$   \\
   \ $|z\>$ &\ $\mu_l=\sigma$,  $\tau_p^z\nu_{pq}^z\tau_q^z =+1$   \\
 \hline
\end{tabular}
\caption{This table describes the correspondence between edge labels in the modified string-net model and the spin configurations in the gauged Toric code. Here $\tau_p$ and $\tau_q$ are the two spins bordering link $l$ and $\nu_{pq}$ is the gauge spin connecting them.}
\label{mapping-table}
\end{table}
\fi

%%%%%%%%%%%%%%%%%%%%%%%%%%%%%%%%
%%%%%%%%%%%%%%%%%%%%%%%%%%%%%%%%
\subsection{Ungauging the doubled Ising string-net model}\label{ungauging}

In this section we show that the symmetric toric code model can be obtained from the doubled Ising string-net model using
a recipe which we will refer to as the `ungauging' procedure. This ungauging procedure is how we originally constructed the symmetric 
toric code model. It is also central to our construction of more general symmetry enriched string-net 
models, as we explain in section \ref{general-framework}.

The ungauging procedure involves modifying both the Hilbert space and Hamiltonian of the doubled Ising string-net model. 
The first step is to enlarge the Hilbert space of the doubled Ising string-net
model by adding a two-state spin $\tau_p$ to every plaquette $p$ of the honeycomb lattice. (The reason that we use 
\emph{two}-state spins is related to the fact that we are ungauging a $\mz_2$ symmetry. For a general finite 
symmetry group $G$, we would add $|G|$-state spins to each plaquette, as we will explain in section \ref{general-construction}). 

The next step is to modify the Hamiltonian of the string-net model (Eq. \ref{doubled-ising-ham}) by adding the term 
$-\sum_l P_l$ where $P_l$ is defined as in (\ref{link-projector}): 
\beq
H\rightarrow H^{di}-\sum_l P_l
\eeq
This term energetically favors states $|\{\tau_p^z, \mu_l\}\>$ in which the domain walls between the $\tau_p^z$ plaquette 
spins coincide with the links $l$ for which $\mu_l = \sigma$. The final step is to replace the $B_{p}^{\sigma}$ term in $H^{di}$ by
\beq\label{bpsigma-replacement}
B_{p}^{\sigma}\rightarrow B^{\sigma}_{p}\tau_p^x
\eeq
This replacement is important because it ensures that $[B_{p}^{\sigma},P_l]=0$. Combining all three steps, we can see 
that the resulting Hamiltonian $H$ is exactly the symmetric toric code model (\ref{toric-code-ham}).

As the name suggests, the ungauging procedure is designed to produce a model that (1) is invariant under a global $\mz_2$ symmetry, namely $S = \prod_p \tau_p^x$, and (2) has the property that if this $\mz_2$ symmetry is gauged, then the resulting gauge theory supports the same types of anyon excitations and braiding statistics as the model we started with, i.e. the double Ising string-net model. In this respect, the ungauging procedure works as advertised since it produces the symmetric toric code which we already know has properties (1) and (2).

At the same time, it is unsatisfying that the above ungauging procedure can only be applied to the doubled Ising model; it would be more natural if we could `ungauge' a large class of models in the same way that we can `gauge' a large class of models. In fact, the ungauging procedure is more general than it appears (though not as general as gauging). For example, we can easily extend the procedure to an arbitrary perturbed doubled Ising Hamiltonian $\tilde{H}^{di}$ such that (a) $\tilde{H}^{di}$ is gapped, (b) $\tilde{H}^{di}$ commutes with $\prod_{l \in v} (-1)^{N_{l \sigma}}$ for every vertex $v$, and (c) the ground state of $\tilde{H}^{di}$ obeys $\prod_{l \in v} (-1)^{N_{l \sigma}} = 1$. To ungauge a Hamiltonian of this kind, we follow the same steps as above except that (\ref{bpsigma-replacement}) needs to be replaced with a more general rule. The more general rule states that any term in the Hamiltonian that flips the sign of $(-1)^{N_{l \sigma}}$ along a closed loop $\gamma$ should be multiplied by $\prod_{p \in \gamma} \tau_p^x$, where the product runs over plaquettes contained within the loop. One can easily check that this more general procedure produces a model that has properties (1) and (2) above, using the same analysis as in section \ref{gauging}. (Of course, another way that the ungauging procedure can be generalized is to consider other symmetry groups and string-net models. We discuss how this works in section \ref{general-framework}).

A few more comments about ungauging: first, we would like to point out that while the ungauging procedure is inverse to gauging at the level of topological properties, it is \emph{not} the strict inverse at a microscopic level. This is clear from the analysis in section \ref{gauging} since gauging the symmetric toric code produces a model which differs slightly from the doubled Ising string-net model (but which nevertheless shares the same topological properties).

Another important point is that the ungauging procedure corresponds to a very simple operation on ground state wave functions. Consider a sphere or infinite plane geometry and let $\Psi^{di}(\{\mu_l\})$ denote the ground state wave function of the doubled Ising string-net model (the input for the procedure) and $\Psi(\{\tau_p^z,\mu_l\})$ denote the ground state wave function of the symmetric toric code model (the output for the procedure). Then the two wave functions are related by
\begin{equation*}
\Psi(\{\tau_p^z, \mu_l\}) =
 \begin{cases}
\Psi^{di}(\{\mu_l\}) \ \text {if $\sigma$'s match $\tau_p^z$ domain walls} \\
0  \ \ \ \ \ \ \ \ \ \ \ \  \text{otherwise}
\end{cases}
\end{equation*}
(See appendix \ref{gs-derivation} for a derivation of this identity). On a torus, the situation is more complicated. In this case the doubled
Ising string-net model has $9$ degenerate ground states, and some of these states have a nonzero amplitude for an odd number of $\sigma$ loops to wrap around at least one of the non-contractible cycles of the torus. In the ungauged model, however, such a state would cost energy since it is not compatible with a domain wall configuration of plaquette spins. This suggests that on a torus the ground state degeneracy, and therefore the topological order, of the two theories is different. This is consistent with the calculation of appendix \ref{degeneracy} which shows that the symmetric toric code has a ground state degeneracy of 4 on a torus. 

%Finally, we mention that the relationship between the doubled Ising string-net and the symmetric toric code can be used to obtain the string operators $W_e$ and $W_m$ from section \ref{toric-code}. The general idea is that the $\sigma\bar{\sigma}$ string operator, $W_{\sigma\bar{\sigma}$ is equal to $W_e+W_m$

%%%%%%%%%%%%%%%%%%%%%%%%%%%%%%%%
%%%%%%%%%%%%%%%%%%%%%%%%%%%%%%%%

\section{General construction}\label{general-framework}

In this section we show that the symmetric toric code can be generalized to a large class of exactly solvable models for SET phases.
These models are very powerful: they can realize \emph{every} symmetry-enriched topological phase whose underlying topological order can be realized by the string-net construction.

Before we get into details, we need to describe the input data that goes into our construction. Suppose we want to build a model for an SET phase with anyon excitations $\mathcal A$. For our construction to work, the collection of anyons $\mathcal A$ must be realizable by a string-net model. Equivalently, in mathematical language,  $\mathcal{A}$ must be the `Drinfeld center' of some `unitary fusion category' $\mathcal C$, which we denote by $\mathcal A= \mathcal Z(\mathcal C)$. In this language, the input data for our construction is a mathematical object called a `$G$-extension' of $\mathcal C$ (we will define these terms below).

To see why $G$-extensions are a sensible choice for input data, it is important to recall a theorem of Ref. \onlinecite{etingof} which states that there is a one-to-one correspondence between $G$-extensions of $\mathcal C$ and braided $G$-crossed extensions of $\mathcal Z(\mathcal C)$.\cite{etingof} As we mentioned in the introduction, the latter objects can be thought of as mathematical descriptions of SET phases with symmetry group $G$ and anyon excitations $\mathcal A = \mathcal{Z}(\mathcal C)$, so this theorem implies that there is a natural correspondence between $G$-extensions and SET phases. Our construction provides a concrete realization of this correspondence.

\begin{figure}[tb]
    \includegraphics[width=.49 \textwidth]{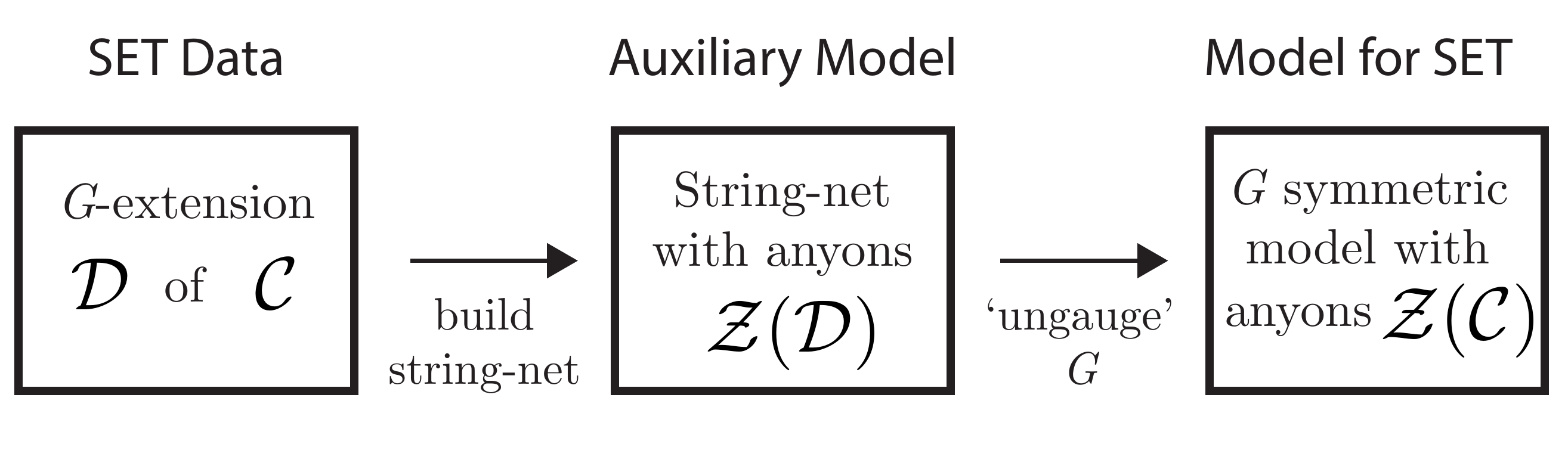}
        \centering
            \caption{Building a symmetry enriched string-net model involves two steps. First we construct the string-net model for
the $G$-extension $\mac{D}$ of $\mac{C}$. Next we `ungauge' the string-net model to obtain a $G$-symmetric model with anyon excitations given by 
$\mac{Z}(\mac{C})$. }
    \label{three_steps}
    \end{figure}

The general idea behind our construction is illustrated in Fig. \ref{three_steps}. Our construction takes as input a $G$-extension $\mathcal D$ of some unitary fusion category $\mathcal C$ and it returns as output an exactly solvable model realizing an SET phase with symmetry group $G$ and anyon excitations $\mathcal A = \mathcal Z(\mathcal C)$. The construction itself proceeds in two steps. The first step is to build the string-net model corresponding to $\mathcal{D}$. This string-net model realizes a topological phase with anyon excitations given by $\mathcal{Z}(\mathcal{D})$. Crucially, it was shown in Refs. \onlinecite{stationq, etingof} that this collection of anyons is precisely what would be obtained if we gauged the global symmetry of the SET phase 
%(i.e. braided $G$-crossed extension of $\mathcal A$)
of interest. In other words, the string-net model realizes the \emph{gauged} SET phase. Hence, to obtain a model for the ungauged SET phase, all we have to do is reverse the gauging procedure --- that is, construct a model that `gauges' into the string-net model based on $\mathcal{D}$. This is the second step illustrated in Fig. \ref{three_steps} and it is accomplished by following a simple recipe which we will describe below.

%The symmetric toric code model is a special case of this construction.

We now briefly review the definitions of the various terms used above and fill in the details of the construction.

%%%%%%%%%%%%%%%%%%%
%\subsection{Mathematical concepts and terminology}\label{review}
%------------------------------------------------
%\subsubsection{Drinfeld center, string-net models, and $G$-extensions}
\subsection{Drinfeld center, string-net models, and $G$-extensions}\label{review}

The \emph{Drinfeld center} construction is a mathematical procedure that takes as input a unitary fusion category $\mathcal{C}$,
and produces, as output, an anyon theory $\mathcal A = \mathcal Z(\mathcal C)$.\cite{Drinfeld} Here, a 
\emph{unitary fusion category} $\mathcal C$ is a finite collection of $N$ simple objects
$i,j,k,\ldots$, that includes a trivial simple object $1$, together with fusion spaces $V^{ij}_k$ and
associativity relations \ {encoded in unitary matrices} $F^{ijk}_l$ satisfying certain coherence conditions. 
An \emph{anyon theory} is also a fusion category, but with the additional data associated with non-degenerate braiding (of anyons).\cite{Kitaev2006}
More physically, an anyon theory is the mathematical data used to describe a topological phase without symmetry.
 
Some simple examples of the Drinfeld center construction are: (1) the toric code, which is the Drinfeld center 
of $\mathcal{C} = \Z_2$, (2) more generally, \ {the irreducible representations of} the quantum double of a finite group $\mac{G}$, which is the Drinfeld center of $\mathcal{C}=\mac{G}$
%[as well as the Drinfeld center of $\mathcal{C} = Rep(H)$, the category of irreducible representations of $H$] 
and (3) any anyon theory of the form 
$\mathcal{T} \times {\overline{\mathcal{T}}}$, which is the Drinfeld center of $\mathcal T$. The Drinfeld center is
sometimes also referred to as the `quantum double' construction, though here we will only use the term quantum double in the context of 
finite groups, as in example (2) above.

\emph{String-net} models are exactly solvable models that provide a physical realization of the Drinfeld center construction.\cite{string-net,Kitaev_Kong} The input 
necessary to build a string-net model is a set of string types, fusion rules and $F$-symbols, obeying certain consistency conditions (see e.g. section \ref{doubled-Ising}). In mathematical language, this input data is precisely a unitary fusion category $\mathcal C$, with string types corresponding to simple objects in $\mathcal C$. Likewise, the output of 
the string-net construction is a lattice spin model that realizes precisely the anyon theory $\mathcal{A} = \mathcal Z(\mathcal C)$.  This model is built out of $|{\mathcal{C}}|$-state spins living on the \emph{links} of the honeycomb lattice. The Hamiltonian takes the form
\begin{equation}
H^{\text{snet}} = - \sum_v Q_v - \sum_p B_p^{\text{snet}}
\label{snet-Ham}
\end{equation}
where the first term, $Q_v$, is a projection operator that projects onto states that obey the fusion rules of $\mathcal C$ at vertex $v$, and the second term, $B_p^{\text{snet}}$, is the string-net plaquette term:
\begin{equation}
B_p^{\text{snet}} = \sum_{s \in \mathcal{C}} a_s B_p^s, \quad \quad a_s = \frac{d_s}{\sum_{i \in\mathcal{C}} d_i^2}
\label{Bpsnet}
\end{equation}
Here $d_s$ denotes the quantum dimension of $s$ and the definition of $B_p^s$ is given in Ref. \onlinecite{string-net}.
%This model consists of 
%spins living on the links of the honeycomb lattice and the Hamiltonian contains terms that enforce fusion rules at vertices, as well as terms that make the 
%string-net configurations fluctuate.\cite{string-net} We will describe this construction in more detail below.

A \emph{$G$-extension} of ${\mathcal{C}}$ is simply another fusion category $\mathcal{D}$ with an additional property called a $G$-grading, \ {which is a} decomposition of $\mathcal{D}$,
\begin{align}
\mathcal{D}=\bigoplus_{g\in G} {\mathcal{D}_g}_,
\end{align}
satisfying ${\mathcal{D}}_g \times {\mathcal{D}}_h \subset {\mathcal{D}}_{gh}$ and ${\mathcal{C}} = {\mathcal{D}}_1$, where $1$ is the identity of $G$. \ {We will always assume that the $G$-grading is \emph{faithful}, i.e. $\mac{D}_g $ is non-empty for all $g\in G$. }

\ {To get a better feeling for this definition we mention a few examples and make a few comments:}

{\bf{(1)}}  To begin, consider the case where $\mathcal C = \mz_2 = \{1,\psi\}$ and $G = \mz_2$. In this case, the Ising fusion category
${\mathcal{D}} = \{1,\psi,\sigma \}$ provides an example of a $\mz_2$ extension of $\mathcal C$ with a $\mz_2$ grading given by
${\mathcal{D}}_{1}= \{1,\psi \}$ and ${\mathcal{D}}_{-1}=\{\sigma\}$ \ {(Here, the $F$ symbols for $\mathcal D$ are given in Eq. [\ref{F-symbols}])}. In fact, this {is precisely the $G$-extension we used} to construct the symmetric toric code model of section \ref{toric-code}.

{\bf{(2)}} \ {Another class of $G$-extensions which may be more familiar are ordinary group extensions. In this case the category $\mac{C}$ and the extended category $\mathcal{D}$ are both groups, and $\mac{C}$ is a normal subgroup of $\mac{D}$ such that the quotient group $\mac{D}/\mac{C}$ is isomorphic to $G$. The $F$-symbols can be taken to be trivial, i.e. $F \equiv 1$, for both $\mac{C}$ and $\mac{D}$. For example, the dihedral group $\mac{D} = D_{2N}$ is a $\mz_2$ extension of $\mac{C} = \mz_N$. The identity component $\mac{D}_1$ is the $\mz_N$ subgroup consisting of rotations while the other component $\mac{D}_{-1}$ consists of elements that are the composition of a rotation and a reflection.} 
%In this example, the $F$-symbols are valued in the group cohomology group $H^3(D_{2n},U(1))$
% and $\mz_2$ acts on the elements of $\mz_N$ by taking each element to its inverse: $i\rightarrow -i$.

{\bf{(3)}} \ {Note that $G$-extensions should not be confused with \emph{braided} $G$-crossed extensions. A $G$-extension is a fusion category and thus only contains data related to fusion; a braided $G$-crossed extension is a more complicated structure that contains both fusion and braiding data, and that can be thought of as a mathematical description of an SET phase.}

{\bf{(4)}} \ {As we mentioned earlier, it is known that there is a one-to-one correspondence between $G$-extensions of $\mathcal{C}$ and braided $G$-crossed extensions of $\mathcal{Z}(\mac{C})$.\cite{etingof} This correspondence plays a central role in our construction, so it would be useful if we could turn it into an \emph{explicit} algorithm that produces a $G$-extension of $\mathcal C$ given a braided $G$-crossed extension of $\mathcal{Z}(\mac{C})$. We do not know of such an algorithm, but there are some partial results in this direction that are worth mentioning. Specifically, one such result is that if a braided $G$-crossed extension has the property that the symmetry action $\rho$ on the anyons is trivial, then the corresponding $G$-extension $\mathcal D$ is always made up of $|G|$ copies of ${\mathcal{C}}$, i.e. ${\mathcal{D}}_g = {\mathcal{C}}$ for all $g$.\cite{etingof}} 
%(Note however that the fusion rules between ${\mathcal{D}}_g$ and ${\mathcal{D}}_h$ do not have to be the usual fusion rules of $\mathcal{C}$ for $g,h \neq 1$).
This also implies that if the $G$-extension $\mathcal D$ is \emph{not} made of $|G|$ copies of $\mathcal{C}$ then the action of $\rho$ is non-trivial. 

For instance, in the example $\mathcal D = \{1, \psi, \sigma\}$ discussed above, the two components of the $G$-extension are $\mathcal{D}_1=\{1,\psi\}$ and $\mathcal{D}_{-1}=\{\sigma\}$. The fact that these two components have different size, and are therefore not isomorphic, immediately implies that in the corresponding SET phase, the symmetry has a non-trivial $\mz_2$ action $\rho$ on the topological data characterizing the phase. In general, such a non-trivial action means that either $\rho$ permutes the anyon types of the SET phase or $\rho$ acts non-trivially on the fusion spaces $V^{ij}_k$ and leaves the anyons fixed. In this example, we know that $\rho$ permutes the anyon types because the SET phase corresponding to this $G$-extension is the symmetric toric code from section \ref{toric-code}.

\begin{comment}
\ {
{\bf{(5)}} Let $D^2$ be the total quantum dimension for a given anyone theory i.e. }
\end{comment}

%%%%%

\subsection{Symmetry enriched string-net models}\label{general-construction}

%\subsubsection{Hilbert space and symmetry action for $G$-symmetric model}

We are now ready to present the general symmetry enriched string-net construction. The input data for our construction is a $G$-extension ${\mathcal{D}}$ of a unitary fusion category ${\mathcal{C}}$. The output is a commuting projector model for an SET phase with symmetry group $G$ and anyon excitations $\mathcal A = \mathcal Z(\mathcal C)$.

As mentioned earlier, our construction has two steps. The first step is to build a string-net model with string types corresponding to the simple objects in $\mac{D}$. This model is built out of $|{\mathcal{D}}|$-state spins living on the links of the honeycomb lattice with a Hamiltonian, $H^{\text{snet}}$, given by Eq. (\ref{snet-Ham}) but with $\mathcal{C}$ replaced by $\mathcal{D}$. The second step is to `ungauge' this string-net model. We do this by modifying both the string-net Hilbert space and the string-net Hamiltonian. Starting with the Hilbert space, the only modification is that we add $|G|$-state degrees of freedom to the \emph{plaquettes} of the honeycomb lattice. The Hilbert space of our model now consists of two separate tensor factors. One tensor factor is formed by the $|G|$-dimensional Hilbert spaces living on the plaquettes $p$, each with basis states $|g_p\>, g_p\in G$, which we refer to as plaquette spins. The other tensor factor is the $|{\mathcal{D}}|$-dimensional string-net Hilbert space living on each link $l$ of the honeycomb lattice with basis states $|s_l\>$, $s_l \in \mathcal{D}$. In this notation, the basis states for the full Hilbert space are labeled as $|\{g_p, s_l\}\>$.

As for the Hamiltonian, we make two modifications to $H^{\text{snet}}$  (Eq. \ref{snet-Ham}). First, we add a new term $-\sum_l P_l$. Second, we modify the string-net plaquette term in a particular manner: $B_p^{\text{snet}} \rightarrow B_p$. The end result takes the form
\begin{align}
H = - \sum_l P_l-\sum_v Q_v  - \sum_p B_p
 \label{gen_ham}
\end{align}
where the sums run over the links ($l$), vertices ($v$), and plaquettes ($p$) of the honeycomb lattice.

In order to explain the three terms listed above, we need to introduce some notation. Our first piece of notation involves the $G$-grading of $\mac{D}$: for each element $s \in \mac{D}$, we denote its corresponding group element $g $ in $G$ by $g_s$. By definition, $s \in \mathcal{D}_{g_s}$. 

\begin{figure}[tb]
    \includegraphics[width=.3\textwidth]{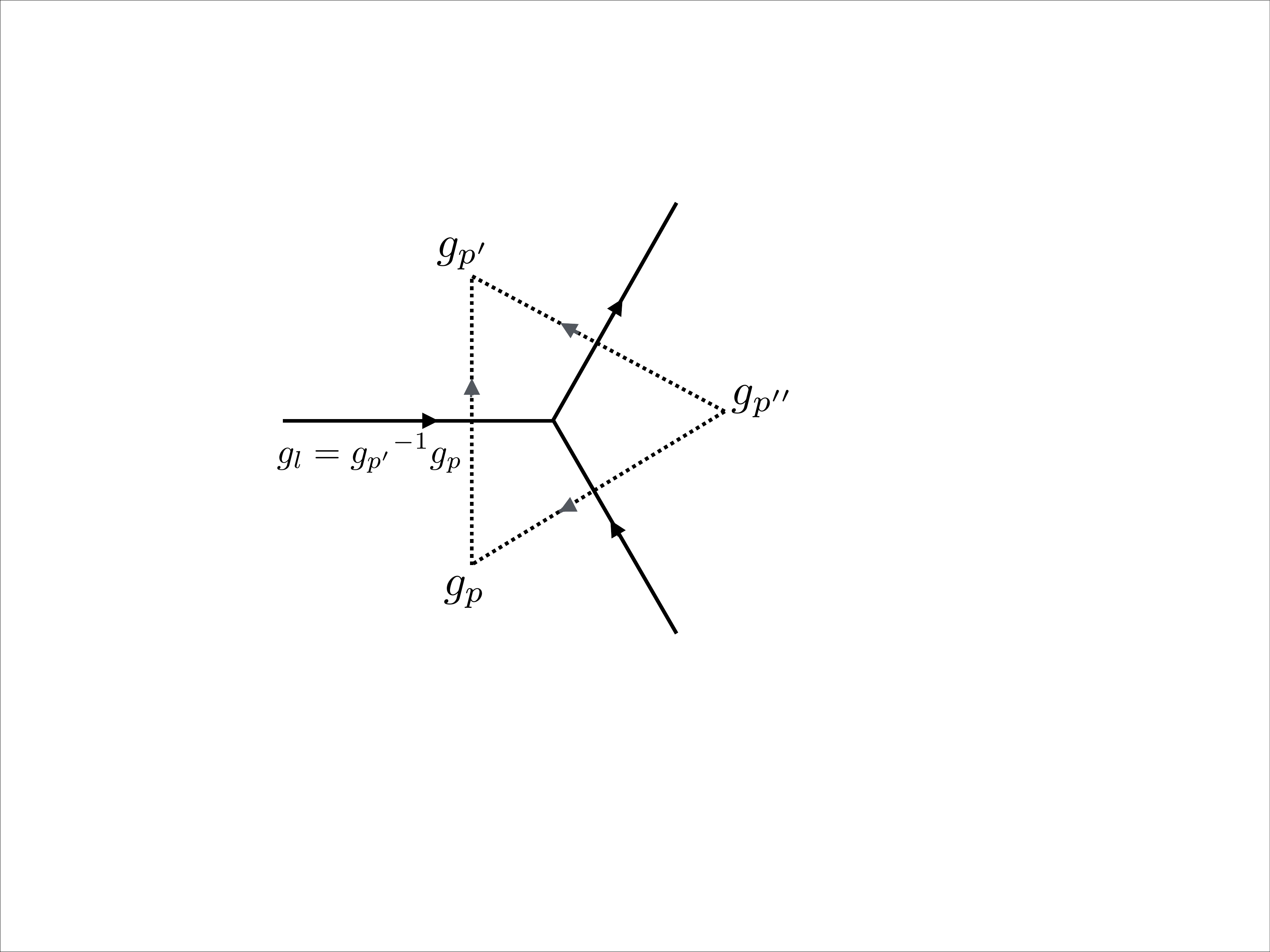}
        \centering
            \caption{By rotating oriented links 90 degrees counterclockwise, an orientation on the original hexagonal lattice gives an 
orientation on the dual triangular lattice.  The latter allows us to define link variables $g_l$, given any configuration of plaquette spins $\{ g_p \}$ and any link $l$ of the original hexagonal lattice.}
    \label{general_config}
    \end{figure}

Another piece of notation has to do with domain walls between plaquette spins. Given any $G$-spin configuration $\{g_p\}$, we define a $G$-valued link variable $g_l$ by:
\begin{align} \label{g_l}
g_l = g_{p'}^{-1} g_p
\end{align}
where $p$ and $p'$ are the two plaquettes that border link $l$, oriented from $p$ to $p'$ (Figure \ref{general_config}). We can think of $g_l$ as defining a domain wall configuration.
Note that in order to make this definition unambiguous we need to specify a convention for which plaquette corresponds to $p'$ and which plaquette corresponds to $p$. To this end, recall that the links in a string-net model always carry an orientation, which we take to be fixed. By rotating the oriented links 90 degrees counterclockwise, we also obtain a fixed orientation of the dual triangular lattice. This orientation allows us to fix a convention for Eq. (\ref{g_l}), as shown in Figure \ref{general_config}.

With this notation, we are now ready to define the different terms in the Hamiltonian (\ref{gen_ham}). The first term $P_l$ is defined by
\beq
P_l|X\>=\begin{cases}
|X\> \  &\text{if } g_l=g_{s_l} \\
\ 0\ &\text{otherwise}
\end{cases}
\eeq
where $|X\>$ is some basis state $|X\> = |\{g_p, s_l\}\>$, $g_l$ is defined as in Eq. (\ref{g_l}) and $s_l$ is the ${\mathcal{D}}$-label on link $l$. The general interpretation of $P_l$ is as follows: there are two different ways to obtain a $G$-valued string-net configuration on links. One comes from the domain walls of the plaquette spins, and the other comes from examining the $G$-grading of the link variables in the ${\mathcal{D}}$ string-net configuration. The term $-\sum_l P_l$ energetically favors configurations where these two agree.  This is a direct generalization of the operator $-\sum_l P_l$ given in the toric code example which favored states for which plaquette spin domain walls coincided with $\sigma$ strings.

The second term, $Q_v$, is simply the usual string-net vertex operator. On the other hand, the third term, $B_p$, is a 
\emph{modified} version of the usual string-net plaquette operator $B_p^{\text{snet}}$ (Eq. \ref{Bpsnet}). In particular,
\begin{align}
B_p = \sum_{s \in \mathcal{D}} a_s B_p^s {\tilde{U}}^{g_s}_p
\end{align}
where ${\tilde{U}}^g_p$ acts as right-multiplication by $g$ 
\begin{align}
{\tilde{U}}^g_p :  |g_p\rangle \rightarrow |g_p g \rangle
\end{align}
Note that the only difference between $B_p$ and $B_p^{\text{snet}}$ is the presence of the operator ${\tilde{U}}^{g_s}_p$, which acts on the plaquette spins. 

\ {We should mention that the above models only apply to the case where the category $\mathcal D$ does not have any fusion multiplicity (a category has \emph{fusion multiplicity} if one of the fusion spaces $V^{ij}_k$ has dimension two or greater). If we want to accommodate fusion multiplicity in $\mathcal D$, we need to modify the corresponding string-net model $H^{\text{snet}}$ by including additional spins that live on the \emph{sites} of the honeycomb lattice.\cite{string-net, Kitaev_Kong, Lan_Wen} Likewise, these site spins also need to be included in the symmetry-enriched string-net model (\ref{gen_ham}). This extension is straightfoward so we omit the details; the main point is that our construction can be applied to arbitrary fusion categories $\mathcal{D}$ as long as one uses suitably generalized string-net models.\cite{Kitaev_Kong, Lan_Wen, Lin_Levin}}

\subsection{Properties}
The Hamiltonian $H$ (\ref{gen_ham}) defined above has a number of interesting properties. For the most part, the proofs of these properties are very similar to the ones given in detail for the symmetric toric code so we do not repeat the analysis here, but rather simply sketch the main idea. 

{\bf{(1)}} $H$ has a global $G$ symmetry, where the symmetry action $U_g$ is defined by left multiplication by $g$:
\begin{align} \label{gen_sym}
U^g &= \bigotimes_p U^g_p \\
U^g_p &: |g_p\rangle \rightarrow |g g_p\rangle
\end{align}
Note that $U_g$ is the natural generalization of the $\mz_2$ symmetry $S = \prod_p \tau^x_p$ in the symmetric toric code model.

{\bf{(2)}} All of the terms in $H$ commute. The proof of this statement is nearly identical to the one given for the symmetric toric code in section \ref{hamiltonian-properties}.
\begin{comment}
 Indeed, the fact that $Q_v$ commutes with $B_p$ follows from the commuting nature of the standard string-net Hamiltonian, and the fact that ${\tilde{U}}^{g_s}_p$ acts only on the $G$-spin degrees of freedom.  Also, $Q_v$ clearly commutes with $P_l$, as both are diagonal in the spin configuration / string-net labeling basis.  The only non-trivial thing to check is that $B_p$ commutes with $P_l$, with $l$ a boundary link of $p$, and that two neighboring $B_p$'s commute.  The first property is true because $B_p$ changes both $g_l$ and $g_{i_l}$ by the same amount.  The second follows from the commuting nature of the original string-net plaquette terms, together with the fact that the ${\tilde{U}}^{g_s}_p$ operators on different plaquettes commute.
\end{comment}

{\bf{(3)}} All of the terms in $H$ are projectors. This is clearly the case for $P_l$ and $Q_v$ which were defined as projectors. One can see $B_p$ is a projector by using Eq. (\ref{projector-identity}) and following the same reasoning as in the symmetric toric code case. 

{\bf{(4)}} In an infinite plane or sphere geometry, the operators $P_l, Q_v, B_p$ have a unique simultaneous eigenstate with eigenvalue $1$. This state is the unique ground state of $H$. The existence and uniqueness of this state can be deduced from the existence and uniqueness of string-net ground states in a sphere geometry, together with the fact that gauging $H$ gives a modified string-net model, as discussed in {\bf (6)}. (Note that this result implies that $H$ does not spontaneously break the symmetry, since it has a \emph{unique} ground state).

{\bf{(5)}} In a sphere or infinite plane geometry, the ground state amplitude of $X = \{g_p, s_l\}$ is given by
\begin{equation}
\Psi(\{g_p,s_l\}) = \begin{cases}
\Psi^{\text{snet}}(\{s_l\}) \ & \text{if $g_l = g_{s_l}$}  \nonumber \\
0 \ & \text{otherwise}
\end{cases}
\label{gswfgen}
\end{equation}
where $\Psi^{\text{snet}}(\{s_l\})$ is the ground state amplitude of $\{s_l\}$ in the string-net model based on ${\mathcal{D}}$.

{\bf{(6)}}  Gauging the global symmetry (\ref{gen_sym}) of this model results in a topological phase described by the anyon theory $\mathcal{Z}(\mathcal{D})$. 
This can be seen by introducing $G$-valued variables on the links of the dual triangular lattice and performing a minimal coupling procedure as described for the symmetric toric code \ {in section \ref{gauging}.} As in that case, it is straightforward to see that the resulting gauged model \ {is unitarily equivalent to a \emph{modified} string-net model that realizes $\mathcal{Z}(\mathcal D)$. This modified string-net model contains a $|G|$ dimensional spin, $\xi_l$, on the links in addition to the usual $\mathcal{D}$-state spin. While the modified model looks different from the usual string-net model, it belongs to the same phase since its ground state is simply the usual string-net ground state tensored with a product state where all of the auxiliary spins $\xi_l$ are in the $|g=1\>$ state. See section \ref{mod-string-net} for a precise definition of this modified string-net model and the associated unitary equivalence, in the case of the symmetric toric code.}

{\bf{(7)}} The Hamiltonian $H$ realizes an SET phase with symmetry group $G$ and anyon excitations $\mathcal A = \mathcal{Z}({\mathcal{C}})$. Furthermore, this SET phase is precisely the one associated with $\mathcal D$ under the correspondence discussed in section \ref{review}. This property follows from property {\bf{(6)}} together with two mathematical results. The first result is that the anyon theory $\mathcal{Z}(\mathcal{D})$ is exactly what would be obtained by gauging the symmetry of the desired SET phase. The second result is that the anyon theory corresponding to a gauged SET phase uniquely determines the original SET phase. \cite{stationq, etingof}

\section{Examples}\label{examples}

In this section we illustrate the general construction with some additional examples. We start with two simple examples: 
a bosonic SPT phase with a $\mz_2$ symmetry and a toric code phase with a $\mz_2$ symmetry that does not permute any anyons. We note that solvable models for both of these phases have been written down previously.\cite{levin2012braiding,chen2013symmetry, Hermele1}
We then discuss two SET phases that have non-abelian anyons and anyon-permuting symmetries --- one with a $\mz_2$ symmetry 
and one with a $\mz_3$ symmetry. 

%The purpose of this section is to both discuss relevant examples and to further illustrate the general framework. 
%To this end, the first two examples we discuss are not new, but they do illustrate the general construction and show how it relates to previous work. We then describe two examples of lattice models with anyonic symmetry: one with a $\mz_2$ symmetry and one with a $\mz_3$ symmetry.  In each of these cases the symmetry permutes anyons which have non-trivial mutual statistics.  

%and begin with two models that do not contain anyonic symmetry.  The first provides a concrete illustration of the relation between our general construction and existing models of SPT phases\cite{LevinGu,Chen2013}; the second gives an example of how our construction applies to models with non-trivial topological order but no anyonic symmetry \cite{Hermele}.

%are several possible choices of $H^2(G, \mac{A})$ and $H^3(G, U(1))$, and hence potentially several distinct models based on the same anyonic symmetry.   Though in principle our approach can realize all of these, we do not attempt to construct them all here.   Rather, in each case, we present one possible choice for which we happen to know the appropriate $G$-extension  $\mac{D}$ of $\mac{C}$; 
%the other distinct phases could in principle be constructed in exactly the same manner, by substituting the appropriate $G$-extension.  
%, corresponding to the case where the $G$-extension $\mac{D}$ of $\mac{C}$ is itself a unitary modular tensor category.

%----------
\subsection{Bosonic $\mz_2$ SPT phase} \label{SPTSubsection}
We begin by constructing a model for a bosonic SPT phase with $\mz_2$ symmetry. The first step of our construction is to find the fusion category $\mathcal C$ associated with this phase. 
To this end we note that, like all SPT phases, this phase supports only trivial anyons: $\mathcal A = \{1\}$. Trivial anyons 
can be realized by a trivial string-net model with only one string type, so we conclude that $\mathcal C = \{1\}$. 
%Another way to say this, in more mathematical language is that $\mathcal %A = \{1\}$ is the Drinfeld center of the trivial fusion category, i.e. %$\mathcal A = \mathcal Z(\mathcal C)$ where $\mathcal C =\{1\}$.

Having found $\mathcal C$, the second step is to find the different $\mz_2$-extensions $\mathcal D$ of $\mathcal C$; each extension
will give us a model for a different $\mz_2$ SPT phase. 
Conveniently the $\mz_2$ extensions of $\mathcal C = \{1\}$ are already known: there are two such extensions, both of which are of
the form $\mathcal D = \mz_2 = \{1,a\}$ with
\begin{align}
a \times a = 1, \quad a \times 1 = a, \quad 1 \times 1 = 1
\end{align}
and with the $\mz_2$ grading given by
$\mathcal{D}=\mathcal{D}_1\oplus\mathcal{D}_{-1}$ with $\mathcal{D}_1 = \{1\}$ and $\mathcal{D}_{-1}= \{a\}$.
\footnote{More generally, for any finite group $G$, the number of 
$G$-extensions of $\{1\}$ is given by $H^3(G,U(1))$ and all the extensions have fusion rules that are isomorphic to $G$.} 
The difference between the two extensions comes from their $F$-symbols $F^{ijk}_l$. In one extension, the $F$ symbol is trivial: 
$F \equiv 1$. In the other extension, $F^{aaa}_a = -1$, and $F = 1$ otherwise. 

Let us focus on the second $\mz_2$ extension, since this is the one that gives the \emph{non-trivial} SPT phase.
To build a model for this phase, we first need to construct the string-net model with string types given by $\mathcal D$. This string-net 
model is well-known and is called the `doubled semion' model.\cite{string-net} We then need to ungauge the doubled semion model. Following the general 
recipe, we add two-state spins to the plaquettes of the honeycomb lattice, with states labeled by $\tau_p^z=\pm$. We then modify the 
Hamiltonian of the doubled semion model by adding the term $-\sum_l P_l$ with
\beq
P_l=\frac{1}{2}(1+(-1)^{N_l}\tau_p^z\tau_q^z)\non
\eeq
Here $p,q$ denote the two plaquettes that adjoin $l$ while the $N_l$ operator is defined by 
$N_l=1$ if the link $l$ is occupied with a type $a$ string and $N_l =0$ otherwise. 
%This term favors states in which the domain walls of the plaquette spins coincide with occupied ($N_l = 1$) links.
Finally, we alter the plaquette operator 
\beq
B_p \rightarrow \frac{1}{2}(1+B_p^{a}\tau_p^x)
\eeq
The result is a model for a non-trivial SPT phase with a $\mz_2$ symmetry $S = \prod_p \tau_p^x$. 
%(Note that a similar procedure can be used to 
%construct SPT phases with more general symmetry group $G$).

The above model is closely related to the one written down in Ref. \onlinecite{levin2012braiding} but the Hilbert space is 
slightly different: in the above model there are two-state spins on both the links and plaquettes of the honeycomb lattice, while the 
model from Ref. \onlinecite{levin2012braiding} only has spins on the plaquettes. To understand the relationship between the two models, 
it is useful to think about their ground states. For the model in Ref. \onlinecite{levin2012braiding}, the ground state amplitude for 
a plaquette spin configuration $X$ is given by $\Psi(X) = (-1)^{N(X)}$ where $N$ is the number of domain wall loops in $X$. In 
comparison, the ground state amplitude in the above model is given by the same formula, but there is an additional constraint that the 
occupied links $l$ must lie along the domain walls of the plaquette spins; if this constraint is not satisfied then the amplitude 
$\Psi = 0$. While the two ground state wave functions are different, it is easy to see that there exists a local unitary transformation 
that transforms the ground state of the above model into the ground state of the other model tensored with a product state made up of 
the link spins. Given this connection, \ {we can conclude that} the two models describe the same phase.

%---------
\subsection{Toric code with a non-permuting $\mz_2$ symmetry}

In this section, we construct a model for a toric code phase with a $\mz_2$ symmetry that does \emph{not} permute the anyons. The first step is to find the fusion category $\mathcal C$ corresponding to the toric code. Since the toric code phase can be realized by a 
string-net model with two string types and $\mz_2$ fusion rules, we have $\mathcal C = \mz_2 = \{1,a\}$. As for the $F$ symbol $F^{ijk}_l$, 
this is trivial for the toric code model: $F \equiv 1$.

The second step is to find the $\mz_2$ extensions $\mathcal D$ of $\mathcal C$; each extension will give us a model for a different
SET phase. In addition to the two $\mz_2$-extensions with $e\leftrightarrow m$ symmetry discussed in section \ref{toric-code} and appendix \ref{other-toric-code}, we expect four $\mz_2$-extensions with a non-permuting $\mz_2$ symmetry. \cite{Lu_Vishwanath, stationq} These four extensions can be found by inspection: two are 
of the form $\mathcal D = \mz_2 \times \mz_2$, and two are of the form $\mathcal D = \mz_4$. Within each pair, the two extensions 
are distinguished from one another by their (extended) $F$-symbols, $F^{ijk}_l$, which we omit here for the sake of brevity.

The first two $\mz_2$ extensions give models that describe either the trivial toric code, or a trivial toric code stacked on top of a 
$\mz_2$ SPT. The other two extensions give models where either \ {the $e$ particle and the $em$ particle carry a half $\Z_2$-charge or the $e$ particle and the $m$ particle carry a half $\Z_2$-charge (the phase where $m$ and $em$ carry a half charge} can be obtained by re-labeling, and does not constitute a distinct SET). 
In all four cases, the $\mz_2$ symmetry does not permute the anyons.
%We note  that in the latter two cases, stacking with a $\mz_2$ SPT does not lead to any new SET phases or equivalently, $\mz_2$ 
%extensions.\cite{Lu_Vishwanath, stationq}

Let us focus on the $\mz_2$ extension corresponding to an SET where \ {the $e$ and the $em$ particles} carry half a charge. In this case,
$\mathcal D = \mz_4 = \{1,a,a^2,a^3\}$ and the $\mz_2$ grading is given by $\mathcal{D}=\mathcal{D}_1\oplus\mathcal{D}_{-1}$ with 
$\mathcal{D}_1=\{1,a^2\}$ and $\mathcal{D}_{-1}=\{a,a^3\}$. The $F$-symbol is trivial for this extension: $F \equiv 1$.
To build a model corresponding to this phase, we need to construct the string-net model with string types given by $\mathcal D$. This 
model is exactly the Kitaev quantum double model\cite{kitaev2003fault} with $G = \mz_4$, defined on the honeycomb lattice.

After building this string-net model, we then need to ungauge it. Following the same procedure as before, we do this by 
attaching additional two-state spins to the plaquettes, with states labeled by $\tau_p^z= \pm$. Next, we add the term $-\sum_l P_l$, with
\beq
P_l=\frac{1}{2}(1+(-1)^{N_l}\tau_p^z\tau_q^z)\non
\eeq
Here $p,q$ denote the two plaquettes that adjoin $l$ while $N_l$ is defined by 
$N_l=1$ if the link $l$ is occupied by $a$ or $a^3$ and $N_l = 0$ otherwise. 
%This term makes it energetically favorable for $\tau_p^z$ domain walls to coincide with 
%links with labels in $\mathcal{D}_{-1}$. 
The last step is to modify the plaquette operators so that they commute with $P_l$:
\beq
B_p\rightarrow \frac{1}{4}(B_p^1+B_p^a\tau_p^x+B_p^{a^2}+B_p^{a^3}\tau_p^x)\non
\eeq
The resulting model has a $\mz_2$ symmetry $S = \prod_p \tau_p^x$ and describes a toric code phase in which the $e$ \ { and the $em$} particles carry a 
half $\mz_2$-charge.

%---------

\begin{table}[tb] % -- In this table we use another labeling, inspired by the organization of the category's anyons given in Station$
\begin{tabular}{|c|c|c|c|c|c|}
\hline
Anyon & topological spin & $d$ & Anyon & topological spin & $d$  \\
\hline
$a\equiv1$ & $0$ & $1$ & $e$ & $\pi$ & $3$ \\
$b$ & $0$ & $1$ &                $f $ & $0$ & $2$ \\
$c$ & $0$ & $2$ &               $g$ & $2 \pi/3$ & 2 \\
$d$ & $0$ & $3$&                $h$& $4 \pi/3$ & 2 \\
\hline
\end{tabular}
\caption{\label{TabDs3} Anyon types and corresponding topological data for the quantum double $\mathcal{Z}(S_3)$.}
\label{s3-table}
\end{table}

\subsection{ Quantum double of $S_3$}

In this section, we consider an SET phase that has the same anyon types as the quantum double of $S_3$ (see Table \ref{TabDs3}), 
and has a $\mz_2$ symmetry that exchanges two of the anyons (namely the $c$ and $f$ anyons). The possibility of such an SET phase 
was discussed previously in Ref. \onlinecite{stationq}; here we construct an exactly solvable model that realizes it.

As always, the first step in our construction is to find the fusion category $\mathcal C$ associated 
with this topological phase. To this end, 
we note that the quantum double of $S_3$ can be realized by a string-net model with three different string types corresponding to
the three irreducible representations of $S_3$.\footnote{Actually, the quantum double of $S_3$ can also be realized by a 
string-net model with string types corresponding to group \emph{elements} $g \in S_3$. Hence, we could equally well take $\mac{C} = S_3$.} 
Labeling these representations by $1,a_1,a_2$ where $1$ is the trivial representation, $a_1$ is the two dimensional representation, 
and $a_2$ is the one-dimensional representation, we deduce that $\mac{C} = \{1,a_1,a_2\}$. The fusion rules can be read off from
the representation theory of $S_3$:
\be
 a_2 \times a_2 = 1,\  a_1 \times a_2 = a_1,\  a_1 \times a_1 = 1 + a_1 + a_2.\non
\ee
Likewise, the $F$-symbol is given by the $6j$-symbol for $S_3$. 

The second step is to find the $\mz_2$ extensions $\mathcal D$ of $\mathcal C$. Each extension defines a model for a different SET phase, so
if we wanted to be systematic, we would find all such extensions. Here we will be less 
ambitious and will simply discuss one example: $\mathcal{D} = SU(2)_4$.
This extension is particularly interesting because, as we will argue below, 
it corresponds to an SET phase where the $\mz_2$ symmetry exchanges $c$ and $f$. 
 
Before proceeding further we need to explain our notation: $SU(2)_4$ denotes the fusion category associated with non-Abelian Chern-Simons theory with gauge group 
$SU(2)$ at level $4$. This fusion category has five simple objects, which can be labeled according to their `spin' as:
$\mathcal{D} = \{ 1, a_{\frac{1}{2}}, a_1,a_{ \frac{3}{2}}, a_{2} \}$. The fusion rules are given by Eq. (\ref{general-fusion}). 
The $\mz_2$ grading is given by  $\mac{D}=\mac{D}_1\oplus\mac{D}_{-1}$ where 
\be\label{s3-extension}
\mac{D}_1 = \mac{C} = \{ 1, a_1, a_{2} \} \ , \ \ \ \mac{D}_{-1} = \{ a_{\frac{1}{2}},a_{ \frac{3}{2}} \}.
\ee
The $F$-symbol is known but we will not reprint it here.\cite{BondersonThesis} 
%We can verify this $\mz_2$ grading by noting that, like the rules for addition of angular momenta, the fusion rules for 
%$SU(2)_4$ are such that any two elements of $\mac{D}_{-1}$ (which carry half-integer `spin') always fuse to give something with
%integer spin which is in $\mac{D}_1$.

To build a model for the corresponding SET phase, we need to construct a string-net model with string types given by the simple objects in 
$\mac{D}$. We then need to ungauge this string-net model. Following the same recipe as before, we do this by
adding two-state spins to each plaquette, with states labeled by $\tau_p^z=\pm$. Next we modify the string-net 
Hamiltonian by adding the term $-\sum_l P_l$ with
\beq
P_l=\frac{1}{2}(1+(-1)^{N_l}\tau_p^z\tau_q^z)\non
\eeq
Here $p,q$ denote the plaquettes that adjoin $l$ while $N_l$ is defined by $N_l=1$ if the edge is occupied by $a_{\frac{1}{2}}$ or 
$a_{\frac{3}{2}}$ and $N_l = 0$ otherwise. Finally we modify the plaquette operator so that it commutes with $P_l$:
\be
B_p \rightarrow \frac{1}{\sum_i d_i^2} \left(  \sum_{s = 0, 1,2} d_s B_p^s + \tau_p^x \sum_{s =\frac{1}{2}, \frac{3}{2}} d_s B_p^s \right )
\ee
The resulting model has a $\mz_2$ symmetry $S = \prod_p \tau_p^x$ and supports the same types of anyons as the quantum 
double of $S_3$. 

To determine how the symmetry acts on these anyons, notice that the two components in Eq. (\ref{s3-extension}) have different sizes, so by
the general result discussed in section \ref{review}, the $\mz_2$ action on
the topological data must be nontrivial. This means that the $\mz_2$ symmetry either (1) permutes some of the anyons or (2) leaves the anyons
fixed but has a nontrivial action on the anyon fusion spaces. Assuming the latter possibility doesn't occur, 
we deduce that the symmetry must exchange $c$ and $f$ since this is the only anyon permutation that is consistent with the data in Table \ref{s3-table}.

\subsubsection{ $\mac{D}=$SU(2)$_{k}$ for $k$ even}

It is worth noting that the above example is part of a larger family of SET phases, all of which have a 
$\mz_2$ anyon-permuting symmetry. These phases correspond to $\mz_2$-extensions of the form $\mac{D} = SU(2)_k$ where $k$ is even
and where $\mac{C}$ is the subcategory of $\mathcal{D}$ consisting of objects with integer `spin.' Here, the objects in $SU(2)_k$ can be
labeled as $\{a_0\equiv 1, a_{1/2},\dots,a_{k/2}\}$. The fusion rules are\cite{BondersonThesis}
\be\label{general-fusion}
a_{l} \times a_{m} = \sum_{n = |l-m|}^{\text{min}(l+m, k - l - m) } a_{n}
\ee
and admit a natural $\mz_2$-grading:
\be
\mac{D}_{1} = \mac{C} = \{ a_l | l \in \mz \} \ , \ \ \ \mac{D}_{-1} =  \{ a_l | l \in \mz + 1/2 \}
\ee
%$\mac{C}$, the set of integer spin anyons, is sometimes referred to as SO(3)$_k$, though in fact it does not correspond to the rational conformal field theory SO(3)$_k$, and for even $k$ is not a UMTC.  
Following the same reasoning as above, we know that the $\mz_2$ symmetry permutes the anyons 
(or at least acts nontrivially on the topological data) in these phases since the two components of $\mac{D}$ have different sizes. Indeed, 
$\mac{D}_1 = \{ a_0, a_1, \dots, a_{k/2} \}$ contains $k/2 + 1$ elements, 
while $\mac{D}_{-1} = \{ a_{1/2}, a_{3/2}, \dots, a_{ (k-1)/2}\}$ contains only $k/2 -1$ elements. %As an aside, we point out that this analysis implies there is no anyon permuting symmetry when $k$ is odd. To see this, we simply note that when $k$ is odd the two components of the $G$-extension would be the same size. 

%Hence, by the general result
%form section \ref{review}, we know that the $\mz_2$ action on the topological data must be nontrivial. 
\iffalse
\begin{table}[tb]
\begin{tabular}{|c|c|c|c|c|c|}
\hline
Anyon & topological spin & $d$ & Anyon & topological spin & $d$  \\
\hline
1 & $0$ & $1$ &$c_1$& $0$ & $3$ \\
$b$ & $2 \pi $& $1$ & $c_2$ & $0$ & $3$ \\
$b^2$ & $2 \pi $& $1$ & $c_3$ & $0$ & $3$ \\
$a$ & $ \pi$ & $3$ &
$a^*$ & $\pi$ & $3$ \\
$X_1$ & 0 & $4$ & $Y_1$ & 0 & $4$ \\
$X_2 $ & $2\pi/3$ & $4$ & $Y_2 $ & $4\pi/3$ & $4$\\
$X_3 $ & $4\pi/3$ & $4$ & $Y_3 $ & $2\pi/3$ & $4$\\
\hline
\end{tabular}
\caption{\label{A4Tab} Anyon types and corresponding topological data for the quantum double $\mathcal{Z}(A_4)$.}
\end{table}

\fi

\begin{table}[tb] % same table as previous one, but with new notation
\begin{tabular}{|c|c|c|c|c|c|}
\hline
Anyon & topological spin & $d$ & Anyon & topological spin & $d$  \\
\hline
1 & $0$ & $1$ &$c_1$& $0$ & $3$ \\
$b$ & $0 $& $1$ & $c_2$ & $0$ & $3$ \\
$b^2$ & $0 $& $1$ & $c_3$ & $\pi$ & $3$ \\
$a$ & $ 0$ & $3$ &
$c_4$ & $\pi$ & $3$ \\
$x_1$ & 0 & $4$ & $y_1$ & 0 & $4$ \\
$x_2 $ & $2\pi/3$ & $4$ & $y_2 $ & $2\pi/3$ & $4$\\
$x_3 $ & $4\pi/3$ & $4$ &  $y_3 $ & $4\pi/3$ & $4$\\
\hline
\end{tabular}
\caption{\label{A4Tab} Anyon types and corresponding topological data for the quantum double $\mathcal{Z}(A_4)$.}
\end{table}

%------------------------------------------------------------------------------------------
\subsection{ Quantum double of $A_4$}

We now consider an SET phase with the same anyon types as the quantum double of $A_4$ (see Table \ref{A4Tab}) with a $\mz_3$ symmetry that
permutes three of the anyons (\ {namely $a, c_1, c_2$}).

The first step is to find the fusion category $\mac{C}$ associated with this phase. To do this, we note that the quantum double of $A_4$
can be realized by a string-net model with string types corresponding to the four irreducible representations of $A_4$. Labeling these
representations by $\{ 1, b, b^2, a\}$ where $1$ denotes the trivial representation, and $b, b^2$ denote one dimensional representations,
and $a$ denotes a three dimensional representation, we deduce that $\mac{C} = \{ 1, b, b^2, a\}$. The fusion rules can be read off
from the representation theory of $A_4$,
\beq
b \times b^2 = 1, \ b \times a = b^2 \times a = a, \ a\times a= 1+b+b^2 + 2a,
\eeq
while the $F$-symbol is given by the $6j$ symbol for $A_4$.

The next step is to find the $\mz_3$ extensions $\mac{D}$ of $\mac{C}$. Again, rather than finding all such extensions,
we will only discuss one example: $\mac{D} = SU(3)_3$. This is a category with $10$ simple objects: $\{ 1, a,b, b^2, d_1, d_2, d_3, e_1, e_2, e_3 \}$. 
The $\mz_3$ grading is:
\begin{align}\label{a4-extension}
\mac{D}_1 = \mac{C} = \{ 1, a,b, b^2 \} \ , \ \ \ &\mac{D}_\omega = \{d_1, d_2, d_3 \} \non ,\\
& \mac{D}_{\omega^2} = \{e_1, e_2, e_3 \}
\end{align}
where we have labeled the elements of $\mz_3$ as $1, \omega, \omega^2$ with $\omega = e^{ 2 \pi i/3}$. 
The fusion rules are:
\begin{align} \label{SU3Fus}
&d_i \times a = \sum_{j=1}^3 d_j, \ e_i \times a = \sum_{j=1}^3 e_j , \  \ e_i \times b = e_{(i \text{ mod 3})+1 }\non \\
&d_i\times e_j=a+b^{(i-j)\text{mod 3}},\ \ d_i \times b = d_{(i \text{ mod 3})+1 } \non \\
&d_1 \times d_2 =d_3 \times d_3 =  e_2 + e_3, \  d_1 \times d_3 = d_2 \times d_2 = e_1 + e_2 \n
&d_2 \times d_3 =d_1\times d_1 = e_1 + e_3, \ e_1 \times e_2 =e_3 \times e_3 =  d_2 + d_3 \  \n
& e_1 \times e_3 = e_2 \times e_2 = d_1 + d_2, \  e_2 \times e_3 =e_1 \times e_1 = d_1 + d_3 
 \end{align}
We omit the $F$ symbol for brevity.

To build a model for the associated SET phase, we need to construct the string-net model with string labels in $\mac{D}$ 
and fusion rules given by Eq. (\ref{SU3Fus}). We then need to ungauge this string-net model. Following the general procedure,
the first step is to add a \emph{three-state} spin to each plaquette. We denote the basis states for this spin by
$|1\>, |\omega\>, |\omega^2\>$ where $\omega = e^{ 2 \pi i/3}$. We also define two generalized Pauli operators that act on these 
basis states: 
\begin{align}
C_p &= |1\>\<1| + \omega \ |\omega\>\<\omega| + \omega^2 \ |\omega^2\>\<\omega^2| \nonumber \\
S_p &= |\omega\>\<1| + |\omega^2\>\<\omega| + |1\>\<\omega^2|
\end{align}
%a `clock operator' $C_p = |1\>\<1| + \omega |\omega\>\<\omega| + \omega^2 |\omega^2\>\<\omega^2|$ 
%and a `shift operator' $S_p = |\omega\>\<1| + |\omega^2\>\<\omega| + |1\>\<\omega^2|$. 
%In this case there are two types of domain walls: using the orientation convention described by Eq. (\ref{g_l}) we may have 
%$g_l=g^{-1}_{p'} g_p = 1, \omega, \omega^2$. 
Next, we add the term $-\sum_l P_l$ to the string-net Hamiltonian, where
\beq
P_l= \frac{1}{3} \sum_{k=0}^2 \left(\omega^{-N_l} C^{-1}_{p'} C_p \right)^k
\eeq
Here $p,p'$ denote the two plaquettes that adjoin $l$, and we assume the orientation convention shown in Fig. \ref{general_config}. The operator $N_l$ is
defined by $N_l = 0, 1, 2$ depending on whether link $l$ is occupied by a string in $\mac{D}_1, \mac{D}_{\omega}, \mac{D}_{\omega^2}$
respectively. Notice that $P_l$ projects onto states that satisfy $C^{-1}_{p'} C_p = \omega^{N_l}$ --- that is, states
in which the $\mz_3$ domain walls between the plaquette spins coincide with the $\mz_3$ grading of the string types on the links. 
The last step is to modify the string-net plaquette term so that it commutes with $P_l$:
  \be
 B_p \rightarrow  \frac{1}{\sum_s d_s^2} \left(\sum_{j=0}^2  \sum_{s\in \mac{D}_{\omega^j}}d_s S_p^jB_p^s \right )
 \ee
Here $s$ ranges over all labels in $\mac{D}$ and $d_s$ is the associated quantum dimension.

The model obtained from this procedure has a $\mz_3$ symmetry $S = \prod_p S_p$ and supports the same types of anyon excitations as 
the quantum double of $A_4$. Furthermore, by the same reasoning as in the previous two examples, we know that 
the $\mz_3$ symmetry acts non-trivially on the topological data in this model. Presumably 
this means that the symmetry permutes the \ {anyons $a, c_1$ and $c_2$} since this is the only non-trivial permutation that is 
consistent with Table \ref{A4Tab}.

%We can argue that the $\mz_3$ symmetry permutes the anyons $c_1,c_2$ and $c_3$ by making two observations. First, note that the 
%3 components of the $\mz_3$-extension (\ref{a4-extension}) do not all have the same size.  This implies that there is some 
%$\mz_3$ action on the topological data. Next, note that of the possible sets of three anyons listed in Table \ref{A4Tab} 
%the only set that is invariant under a $\mz_3$ symmetry is $c_1$, $c_2$ and $c_3$. 

%%%%%%%%%%%%%%%%%%%%%%%%%%%%%%%
%
%%%%%%%%%%%%%%%%%%%%%%%%%%%%%%%%
\section{Conclusion}

In this paper we have constructed a large class of exactly solvable models for 2D bosonic SET phases with finite, unitary onsite symmetry 
group. These models are very general and can realize every SET phase (or more precisely, every braided $G$-crossed extension) in this class
whose underlying topological order can be realized by a string-net model. As an example of our construction, we have presented an
exactly solvable model with the same anyon excitations as the toric code model and a $\mz_2$ symmetry that permutes the anyons $e$ and $m$.
%We conjecture that our models are as general as possible in the sense that they can realize every SET phase in this class whose 
%underlying topological order (i.e. set of anyon excitations) can be realized by a commuting projector model. 

An interesting corollary of our construction is that it proves that all of the above SET phases are physically realizable in 2D systems. This is 
significant because mere algebraic consistency of the topological data does not guarantee the existence of a physical realization: in principle, there could be additional obstructions to realizing an SET phase in a physical system that are not captured by the known mathematical structure.
Our construction proves that no such obstructions exist, at least in the above cases.
 
One shortcoming of our current construction is that if one wants to construct a model for a specific SET phase, one needs to know the corresponding $G$-extension ahead of time. For example, in the case of the symmetric toric code we had to know that the correct $G$-extension to realize a phase with $e\leftrightarrow m$ symmetry was the Ising category $\mathcal{D}=\{1,\psi,\sigma\}$. In practice, however, 
one is more likely to be faced with a situation where one 
is interested in an SET phase with certain physical properties but one does not have any a priori knowledge of the 
corresponding $G$-extension. In this case, one would first need to translate from the known physical properties to 
the corresponding $G$-extension before applying our 
construction, and this is not necessarily an easy task.

Given this issue it would be desirable to construct models starting from a more convenient set of input data. For 
example, one might want to start from a particular symmetry action on a set of anyons (e.g. $e\leftrightarrow m$), and
construct all possible models that realize this action. This is a challenging problem, but it may be possible
to make progress using previously known mathematical results\cite{etingof} on the relationship between symmetry actions and $G$-extensions.
%One way to solve this problem would be to work out the explicit relationship
%between symmetry actions and $G$-extensions. 
%This relationship may be implicit in the work of Ref. \onlinecite{etingof}, 
%but more work needs to be done to get it into a useful form, which could be the subject of future work.

Apart from this technical issue, perhaps the most interesting direction for future work would be to generalize our construction to a larger class of SET phases. 
For instance, it would be desirable to generalize these models to include continuous and/or anti-unitary symmetries --- especially 
since many physical symmetries of interest fall into these classes. Another 
potentially interesting set of generalizations would be to extend these models to higher dimensional or fermionic systems.

%%%%%%%%%%%%%%%%%%%%%%%%%%%%%%%%

\section*{Acknowledgements}
Near the completion of this paper we learned of a work by Jiang, Cheng, Gu and Qi with related results. \cite{Chengetal}

CH and ML are supported in part by the NSF under grant No. DMR-1254741.
FJB is supported by NSF DMR-1352271 and Sloan FG-2015-65927.  LF is supported by NSF grant
DMR-1519579 and by the Alfred P. Sloan foundation.

%%%%%%%%%%%%%%%%%%%%%%%%%%%%%%%%
%%%%%%%%%%%%%%%%%%%%%%%%%%%%%%%%

\begin{appendix}

%%%%%%%%%%%%%%%%%%%%%%%%%%%%%%%%
%%%%%%%%%%%%%%%%%%%%%%%%%%%%%%%%

\section{Ground state degeneracy}\label{degeneracy}
In this appendix we calculate the degeneracy of the symmetric toric code on a sphere and torus and discuss the generalization to higher 
genus surfaces.\footnote{To define the symmetric toric code on a sphere or on other surfaces, we generalize the model from the honeycomb lattice to an arbitrary trivalent lattice in the obvious way.}  To begin, recall that the ground state degeneracy can be expressed using the following formula:
\beq
D=\text{Tr}\left(\prod_l P_l \prod_v Q_v \prod_p B_p\right)
\eeq
Next, observe that the operator  $\prod_l P_l\prod_v Q_v$ projects onto the space of states that obey the fusion rules at each vertex and whose plaquette spin domain walls coincide with $\sigma$ strings. Denoting this space of states by $\mathcal{X}$, 
we can simplify the calculation by only tracing over this space:
\beq\label{trace-over-X}
D=\text{Tr}_{\mathcal{X}}\left(\prod_p B_p\right)
\eeq
Next, recall that $B_p$ is a sum of three terms: $B_p=\frac{1}{4}(1+B_p^{\psi}+\sqrt{2}B_p^{\sigma}\tau_p^x)$. However, $B_p^{\sigma}\tau_p^x$ is 
clearly traceless because it contains the operator $\tau_p^x$. We can therefore ignore this term in the trace and Eq. (\ref{trace-over-X}) 
simplifies further to
\beq\label{trace-simplified}
D=\text{Tr}_{\mathcal{X}}\left(\prod_p\frac{1}{4}(1+B_p^{\psi})\right)
\eeq

To calculate the trace, it is convenient to use the basis states $|\{\tau_p^z, \mu_l\}\>$. We first note that
\beq\label{tadpole}
\<\text{tadpole}|\prod_{p} \frac{1}{4} (1+ B_p^{\psi})|\text{tadpole}\> =0
\eeq
where $|\text{tadpole}\>$ is any basis state $|\{\tau_p^z, \mu_l\}\>$ in $\mathcal X$ 
containing a $\sigma$ loop with an odd number of $\psi$ strings ending on it (these are states for which $f(X)=0$ in Eq. (\ref{gsform}). We will 
not prove this result here, but it is easy to show using expression (\ref{bppsi}) for $B_p^{\psi}$. 

From (\ref{tadpole}) it follows that the only basis states in $\mathcal X$ that contribute to the trace are the \emph{no-tadpole} states, i.e. states
in which every $\sigma$ loop has an even number of $\psi$ strings ending on it. Denoting this set of no-tadpole states by $\mathcal{Z}$, we
derive
\beq
D = \sum_{Z \in \mathcal{Z}} \<Z| \prod_p\frac{1}{4}(1+B_p^{\psi})|Z\> 
\eeq
Expanding out the product, we obtain
\beq\label{trace-simplified2}
D = \frac{1}{4^N}\sum_{Z \in \mathcal{Z}} \sum_{R} \<Z| \prod_{p \in R} B_p^{\psi}|Z\>
\eeq
where $N$ is the number of plaquettes and the second sum runs over subsets $R$ of the set of plaquettes.
To proceed further, we note that
\beq\label{no-tadpole}
\<Z|\prod_{ p \in R} B_p^{\psi}|Z\> =0\  \text{or } 1
\eeq
for any no-tadpole state $Z \in \mathcal Z$ and any subset of plaquettes $R$. Again, we will not prove this identity here, but it can be derived 
using (\ref{bppsi}). Combining (\ref{trace-simplified2}) and (\ref{no-tadpole}), we can rewrite $D$ as
\beq\label{trace-simplified3}
D = \frac{M}{4^N} 
\eeq
where $M$ is the number of pairs $(|Z\>,R)$ such that $\prod_{ p \in R} B_p^{\psi}|Z\> = |Z\>$.

Our problem is now to compute $M$. Conveniently, this counting problem can be reduced to a simple group theory calculation. The key point is 
that we can think of each operator $\prod_{ p \in R} B_p^{\psi}$ as an element of the group $G =(\mz_2)^N$ since $(B_p^\psi)^2 = 1$, and since the $B_p^\psi$ operators commute with one another. From this point of view, the $B_p^\psi$ operators define an action of the group $G = (\mz_2)^N$ 
on the subpace spanned by the no-tadpole states. In fact, the $B_p^\psi$ operators also define a group action on the \emph{set} $\mathcal Z$ 
(as opposed to the subspace spanned by $\mathcal Z$) since one can easily see that when the operator $B_p^\psi$ acts on a no-tadpole state, it always gives 
back another no-tadpole state (up to a $\pm$ sign, which we will ignore). 

With this identification, the problem of computing $M$ is equivalent to finding the number of fixed points of the above group action for each group
element $g \in (\mz_2)^N$ and then summing over all $g$. Conveniently, the latter quantity can be related to the number of orbits of the group
action via Burnside's lemma:
\begin{align} \label{Mform}
M = |(\mz_2)^N| \cdot (\text{number of orbits of group action}) 
\end{align}

All that remains is to find the number of orbits of the above group action. To this end, we observe that the $B_p^\psi$ operators do not 
affect the $\sigma$ strings or the spins $\tau_p^z$, so there is at least one orbit for each configuration of $\sigma$ loops and $\tau^z_p$ spins.
In fact, in a spherical geometry, it is not hard to see that there is \emph{exactly} one orbit for each configuration of $\sigma$ loops and 
$\tau^z_p$ spins. It follows that the number of orbits is $2^N$ since there are $2^{N-1}$ different configurations of $\sigma$ loops
and there are $2$ possibilities for the $\tau^z$ spins for each $\sigma$ loop confiuration. Substituting this into (\ref{Mform}) and
(\ref{trace-simplified3}), we immediately derive
\beq
D_{\text{sphere}}=\frac{2^N \cdot 2^N }{4^N} =1
\eeq

In contrast, on a torus the situation is different because there are $4$ orbits for each configuration of $\sigma$ loops and
$\tau^z_p$ spins. This additional factor of $4$ comes from the fact that the $B_p^\psi$ operators do not affect the parity of
the number of $\psi$ strings wrapping around the two non-contractible cycles of the torus; the $4$ orbits correspond to the four possibilities
of even/odd parity for each of the two cycles. We conclude that the number of orbits is $4 \cdot 2^{N}$, which gives a degeneracy of
\beq
D_{\text{torus}}=4
\eeq
More generally, on a surface of genus $g$, the number of non-contractible cycles is $2g$, so the number of orbits is $4^{g} \cdot 2^N$
and the degeneracy is $D_g = 4^g$.
%%%%%%%%%%%%%%%%%%%%%%%%%%%%%%%%
%%%%%%%%%%%%%%%%%%%%%%%%%%%%%%%%

 \section{Derivation of ground state wave function}\label{gs-derivation}
In this appendix we explain how to derive formula (\ref{gsform}) for 
the ground state wave function of the symmetric toric code model. The first step is to solve the three 
eigenvalue equations in (\ref{gscond}), each of which tells us something about the structure of the ground state $|\Psi\>$. The first equation, 
$P_l |\Psi\> = |\Psi\>$, implies that the only configurations $|\{\tau^z_p, \mu_l\}\>$ that appear in $|\Psi\>$ are those with 
$\mu_l = \sigma$ along the domain walls of the $\tau^z_p$ spins. The second equation, $Q_v |\Psi\> = |\Psi\>$, tells us that 
the only configurations that appear in $|\Psi\>$ are those in which the $\mu_l$ states obey the fusion rules at every vertex 
(Fig. \ref{branching-rules}). The last equation, $B_p |\Psi\> = |\Psi\>$, is more subtle and tells us something about the 
relative amplitudes of the configurations in the ground state. In fact, this equation tells us that these amplitudes are 
equal to the corresponding ground state amplitudes of the doubled Ising string-net model:
\begin{equation}
\Psi(\{\tau^z_p, \mu_l\}) = \Psi^{di}(\{\mu_l\})
\label{wfconn}
\end{equation}
Here $\Psi^{di}$ denotes the ground state of the doubled Ising string-net model and $|\{\mu_l\}\>$ denotes a string-net basis 
state with string occupations specified by $\mu_l$. 

To prove Eq. (\ref{wfconn}), it suffices to show that the $|\Psi\>$ defined in Eq. (\ref{wfconn}) satisfies $B_p |\Psi\> = |\Psi\>$. 
To this end, we make three observations. The first observation is that the $|\Psi\>$ defined in Eq. (\ref{wfconn}) is symmetric 
under the $\mz_2$ symmetry $S$ so it can be expanded as a linear combination of \emph{symmetrized} states, defined by:
\begin{equation}
|\{\mu_l\},\text{sym}\> \equiv \frac{1}{\sqrt{2}}(|\{\tau^z_p, \mu_l\}\> + |\{-\tau^z_p, \mu_l\}\>)
\end{equation}
Furthermore, Eq. (\ref{wfconn}) tells us that the expansion coefficients in this linear combination are given by
\begin{equation}
\<\{\mu_l\},\text{sym}|\Psi\> = \sqrt{2} \<\{\mu_l\}|\Psi^{di}\>
\end{equation}
The second observation is that the matrix elements of $B_p$ between symmetrized states are identical to the corresponding matrix 
elements of the plaquette operator $B_p^{di}$ in the doubled Ising string-net model:
\begin{equation}
\<\{\mu_l'\},\text{sym}| B_p |\{\mu_l\},\text{sym}\> = \<\{\mu_l'\}| B_p^{di} |\{\mu_l\}\>
\end{equation}
Indeed, this identity follows easily from the definition of $B_p^{di}$ --- namely 
$B_p^{di} \equiv a_1 B_p^1 + a_\psi B_p^\psi + a_\sigma B_p^\sigma$. The third and final observation is that the doubled 
Ising ground state obeys $B_p^{di} |\Psi^{di}\> = |\Psi^{di}\>$. Putting these three observations together, the claim 
follows immediately.

With Eq. (\ref{wfconn}) in hand, all we have to do to derive formula (\ref{gsform}) is to compute the ground state wave function of
the doubled Ising string-net model and show that it agrees with (\ref{gsform}). This computation is discussed in 
section \ref{doubled-Ising}.

%%%%%%%%%%%%%%%%%%%%%%%%%%%%%%%%
%%%%%%%%%%%%%%%%%%%%%%%%%%%%%%%%
\section{The other toric code with $e\leftrightarrow m$ symmetry}\label{other-toric-code}
Previous work has shown that there are \emph{two} distinct toric code phases that have a $\mz_2$ symmetry that
exchanges $e$ and $m$.\cite{stationq} However, we have only discussed how to realize one of these phases --- namely 
the symmetric toric code. In this appendix, we briefly explain how to realize the other phase.

First we need to describe the $\mz_2$-extensions corresponding to the two phases. The $\mz_2$ extension for the 
symmetric toric code is $\mathcal D = \{1,\psi,\sigma\}$ with fusion rules and $F$-symbols coming from the Ising fusion category. 
In comparison, the $\mz_2$ extension for the other toric code phase is also $\mathcal D = \{1,\psi,\sigma\}$ but with $F$-symbols 
coming from the (Ising)$^3$ fusion category.\cite{Kitaev2006} Note that the two extensions are nearly identical to one another, with 
the only difference being that they have slightly different $F$-symbols. 

Because the two extensions are so similar, the corresponding lattice models are also closely related.
Indeed, using the general construction from section \ref{general-framework}, we find that 
the Hamiltonian for the other toric code phase is identical to that of the symmetric toric code except that
the $B_p^{\sigma}$ term given in Eq. (\ref{bpsigma}) is replaced with
\begin{align}\label{bpsigma2}
B_{p,ghijkl}^{\sigma,g'h'i'j'k'l'}(abcdef) = \delta_{\sigma gg'} \cdots \delta_{\sigma ll'} \cdot (-2)^{-\frac{N_{p\sigma}}{4}}  (-1)^{N_{p2}}
\end{align}
(Here, the only change is $2 \rightarrow -2$). This minor change in the Hamiltonian also leads to a minor modification of the
ground state wave function. Instead of Eq. (\ref{gsform}) we have
\beq
\Psi(X) \equiv \<X |\Psi\> = (-\sqrt{2})^{N_{\sigma}(X)} f(X).
\label{gsform2}
\eeq
Everything else, including the form of the string operators, is identical to the symmetric toric code case.

To see that these two models belong to distinct SET phases, consider their corresponding gauge theories. By construction,
gauging the symmetric toric code model gives a system with Ising$ \times\overline{\text{Ising}}$ anyons, while
gauging the other model gives a system with (Ising)$^3 \times(\overline{\text{Ising}})^3$ anyons. Since these
two gauge theories have distinct braiding statistics, it follows that the two ungauged models cannot be smoothly
connected without closing the energy gap or breaking the symmetry. \cite{levin2012braiding}

%%%%%%%%%%%%%%%%%%%%%%%%%%%%%%%%%%%%%%%%
%%%%%%%%%%%%%%%%%%%%%%%%%%%%%%%%%%%%%%%%
\section{Derivation of string operators}\label{string-operators-ungauged}

In this appendix, we derive the $W_e$ and $W_m$ string operators (\ref{w_e}), (\ref{w_m}). We do this in two steps: first we construct string operators for the doubled Ising string-net model, and then we translate these over to the symmetric toric code using the connection between the two models.

To begin, we need to understand the relationship between the anyon excitations in the two systems. Given that the doubled Ising string-net model is equivalent to the \emph{gauged} toric code model, we know that one of the anyon excitations of the former system can be identified with the $\mz_2$ gauge charge in the latter system. The obvious candidate is $\psi \bar{\psi}$ since this is the only abelian excitation with bosonic statistics. With this identification, we immediately deduce that the anyons $\{\sigma, \bar{\sigma}, \sigma \bar{\psi}, \bar{\sigma} \psi\}$ correspond to $\mz_2$ gauge fluxes since they have mutual statistics of $\pi$ with respect to $\psi \bar{\psi}$. Likewise, the remaining anyons, $\{1,\psi,\bar{\psi},\psi\bar{\psi},\sigma\bar{\sigma}\}$, correspond to zero-flux excitations since they have mutual statistics of $0$ with respect to $\psi \bar{\psi}$. 

Consider the latter set of anyons $\{1,\psi,\bar{\psi},\psi\bar{\psi},\sigma\bar{\sigma}\}$ and the string operators $\{W_1, W_\psi, W_{\bar{\psi}},W_{\psi \bar{\psi}}, W_{\sigma \bar{\sigma}}\}$ that create them. Because these anyons do not involve any gauge flux, it follows that they can be `ungauged' and mapped onto excitations of the symmetric toric code model. In terms of string operators, this means we can ungauge $\{W_1, W_\psi, W_{\bar{\psi}},W_{\psi \bar{\psi}}, W_{\sigma \bar{\sigma}}\}$ so as to construct string operators in the symmetric toric code model. In fact, by following the ungauging procedure described in section \ref{ungauging}, one can show that this ungauging step is completely trivial in this case: to ungauge one of the above string operators, we simply embed them, without any changes, in the symmetric toric code Hilbert space. We will therefore abuse notation and use the same symbol $W_a$ to denote a string operator and its ungauged counterpart.

After this ungauging step, $\{W_1, W_\psi, W_{\bar{\psi}},W_{\psi \bar{\psi}}, W_{\sigma \bar{\sigma}}\}$ define string operators in the symmetric toric code model, which means they can be expanded as linear combinations of the four elementary string operators $\{W_1, W_e, W_m, W_{em}\}$. Conveniently, the coefficients in these linear combinations are fixed by general considerations (up to local operators acting at the ends of the strings):
\begin{align}
W_\psi &\sim W_{em}, \quad W_{\bar{\psi}} \sim W_{em}, \quad W_{\psi \bar{\psi}} \sim W_1 \nonumber \\
W_{\sigma \bar{\sigma}} &\sim W_e + W_m \label{eplusm}
\end{align}
Here, the first two relations follow from the fact that $W_\psi$ and $W_{\bar{\psi}}$ create fermions and therefore must map onto $W_{em}$, while the third relation follows from the fact that $\psi \bar{\psi}$ is a $\mz_2$ gauge charge, which is trivial in the ungauged theory. As for the last relation, this follows from two observations: (1) $\sigma \bar{\sigma}$ has mutual statistics $\pi$ with respect to $\psi$, and (2) $W_{\sigma \bar{\sigma}}$ is \emph{even} under the $\mz_2$ symmetry (as is every gauge invariant operator). The first observation implies that $W_{\sigma \bar{\sigma}}$ must be a linear combination of $W_e$ and $W_m$ (since these are the only anyons with mutual statistics $\pi$ with respect to $em$), while the second observation implies that the coefficents of $W_e, W_m$ are equal.

\begin{figure}[tb]
    \includegraphics[width=.38\textwidth]{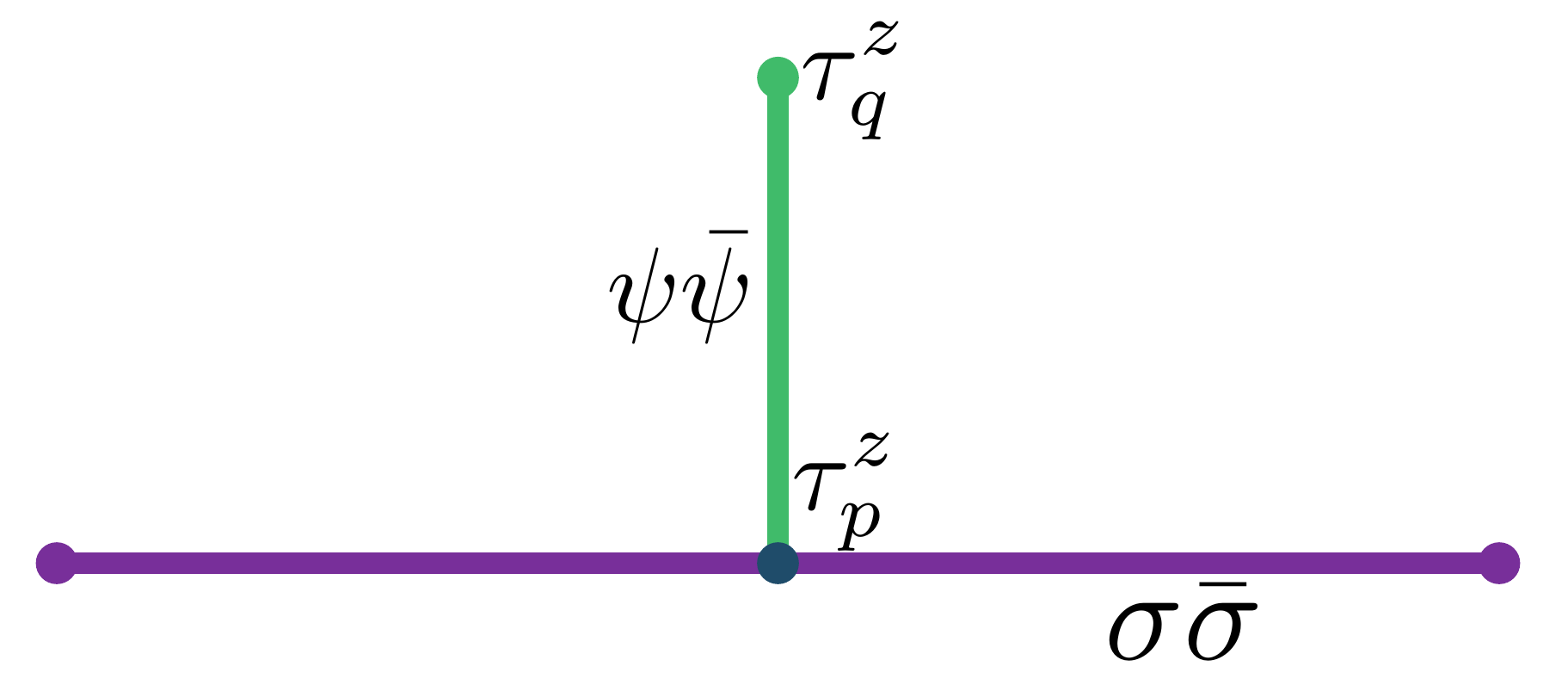}
        \centering
          \caption{The `$T$-junction' operator $W_{\sigma\bar{\sigma}\bot\psi\bar{\psi}}$ is made up of two string operators $W_{\sigma \bar{\sigma}}$, $W_{\psi \bar{\psi}}$ which meet at a point $p$.}
    \label{e-m-string}
    \end{figure}

The last equation (\ref{eplusm}) is especially useful since it provides a route to constructing $W_e + W_m$. This is nice because if we can also find the difference $W_e - W_m$, then we can obtain $W_e$ and $W_m$ individually. The problem is that $W_e - W_m$ is \emph{odd} under the $\mz_2$ symmetry, so it has no counterpart in the doubled Ising string-net model (i.e., it cannot be gauged). To get around this obstacle, we consider the related operator $(W_e - W_m) \tau_q^z$. This operator is \emph{even} under the symmetry so it can be gauged and hence translated into the doubled Ising string-net model. A natural guess is that its gauged counterpart is the `$T$-junction' operator $W_{\sigma\bar{\sigma}\bot\psi\bar{\psi}}$ consisting of a $\psi\bar{\psi}$ string ending on a $\sigma\bar{\sigma}$ string (Fig. \ref{e-m-string}). Thus we have
\begin{align}
\label{eminusm}
W_{\sigma\bar{\sigma}\bot\psi\bar{\psi}} \sim (W_e - W_m) \tau_q^z
\end{align}

Eqs. (\ref{eplusm}) and (\ref{eminusm}) form the backbone of our derivation. The rest is straightforward: we will simply construct the two operators $W_{\sigma \bar{\sigma}}$, and 
$W_{\sigma\bar{\sigma}\bot\psi\bar{\psi}}$ using the general string-net formalism\cite{string-net} and then plug them into (\ref{eplusm}) and (\ref{eminusm}) and solve for $W_e$ and $W_m$. 

We begin with $W_{\sigma \bar{\sigma}}$. This operator is easiest to describe using the graphical representation discussed in Ref. \onlinecite{string-net}. Let $|X\>$ be some Ising string-net state on the honeycomb lattice and let $\gamma$ be a path drawn on the links of this lattice. To compute $W^\gamma_{\sigma\bar{\sigma}} |X\>$, the first step is to shift the path $\gamma$ slightly so that it no longer lies exactly on the honeycomb lattice. The precise way in which $\gamma$ is shifted is not important, but we will use a convention where $\gamma$ is shifted to the `left', where `left' is defined with respect to some arbitrary orientation of $\gamma$. The second step is to insert a `$\sigma\bar{\sigma}$' string along the path $\gamma$ and to resolve all the resulting crossings using the local rules
\begin{align}
  & \left|
  \begin{minipage}[h]{0.15\linewidth}
        \vspace{0pt}
        \includegraphics[height=.3in]{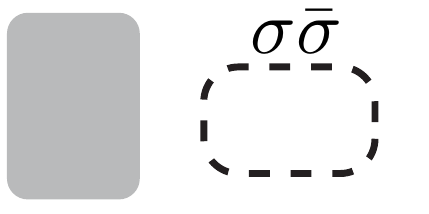}
   \end{minipage}\right\rangle=
 \left|
  \begin{minipage}[h]{0.15\linewidth}
        \vspace{0pt}
        \includegraphics[height=.3in]{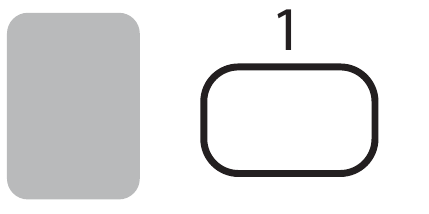}
   \end{minipage}\right\rangle\ + \left|
  \begin{minipage}[h]{0.15\linewidth}
        \vspace{0pt}
        \includegraphics[height=.3in]{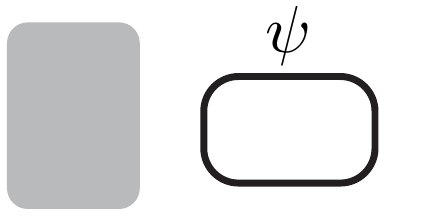}
   \end{minipage}\right\rangle\ 
\label{loop-rule}\\
   &\left|
  \begin{minipage}[h]{0.15\linewidth}
        \vspace{0pt}
        \includegraphics[height=.3in]{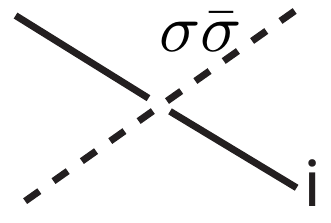}
   \end{minipage}\right\rangle=\sum_{jst}\Omega^j_{sti}
 \left|
  \begin{minipage}[h]{0.12\linewidth}
        \vspace{0pt}
        \includegraphics[height=.3in]{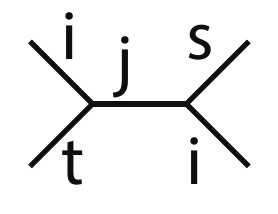}
   \end{minipage}\right\rangle\ 
\label{resolution-rule-over} \\
  &\left|
  \begin{minipage}[h]{0.15\linewidth}
        \vspace{0pt}
        \includegraphics[height=.3in]{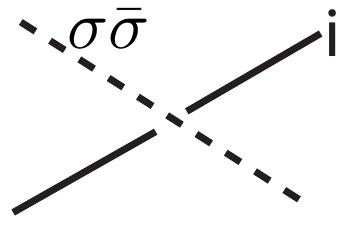}
   \end{minipage}\right\rangle=\sum_{jst}\bar{\Omega}^j_{sti}
 \left|
  \begin{minipage}[h]{0.12\linewidth}
        \vspace{0pt}
        \includegraphics[height=.3in]{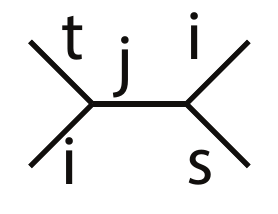}
   \end{minipage}\right\rangle\ 
\label{resolution-rule-over}
\end{align}
Here $\Omega^j_{sti} = \bar{\Omega}^j_{sti}$ is a four index tensor characterizing the string operator whose only nonzero values are
\begin{align}
&\Omega^0_{000}=\Omega^2_{220}=\Omega^0_{222}=\Omega^1_{021}=\Omega^1_{201}=1,\ \ \Omega^2_{002}=-1\non
\end{align}
where $0=\text{vacuum}$, $1=\sigma$ and $2=\psi$. 

After resolving the crossings using the above rules, the result is a linear combination of doubled Ising string-net states, each of which is slightly displaced from the honeycomb lattice. The last step is to `fuse' these strings onto the links of the honeycomb lattice using the rules (\ref{topinv}-\ref{fusion}). Denoting the resulting linear combination of states by $\sum_j c_j |Y_j\>$, the action of $W_{\sigma \bar{\sigma}}$ is then defined by $W_{\sigma \bar{\sigma}}|X\> = \sum_j c_j |Y_j\>$.
 
By examining the above local rules, we can read off the basic structure of the $W_{\sigma \bar{\sigma}}$ string operator. For example, Eq. (\ref{loop-rule}) implies that whenever we resolve the $\sigma\bar{\sigma}$ string, it is always in either the $1$ or $\psi$ channel. Meanwhile, the fact that $\Omega^1_{st1} = 0$ unless $s=0$ and $t=2$ (or vice versa) tells us that $\sigma\bar{\sigma}$ alternates between the $1$ channel and the $\psi$ channel every time it crosses a $\sigma$ string. Putting these two observations together, it follows that $W_{\sigma\bar{\sigma}}$ can be written as a sum of two pieces
    \beq \label{w1w2}
    W_{\sigma\bar{\sigma}}=W_1+W_2
    \eeq
where $W_1$ inserts an alternating $1-\psi-1-\psi \cdots$ string and $W_2$ inserts an alternating $\psi-1-\psi-1 \cdots$ string (with alternations occurring every time the path $\gamma$ crosses a $\sigma$ string).

%The only non-trivial value is $\Omega^2_{002}=-1$ which implies that if the $\sigma\bar{\sigma}$ string crosses a $\psi$ string while in the $1$ channel the state will be multiplied by a $-1$. There can also be additional phases which arise once you fuse the strings onto the lattice. 

Now imagine we ungauge the operator $W_{\sigma \bar{\sigma}}$. As we mentioned earlier, this ungauging process simply amounts to embedding $W_{\sigma \bar{\sigma}}$ within the larger Hilbert space of the symmetric toric code. Although the operator is unchanged, this new context allows us to rewrite $W_{\sigma \bar{\sigma}}$ in a slightly different form. To see this, recall that $W_1$ and $W_2$ alternate between $1$ and $\psi$ each time they cross a $\sigma$ string. At the same time, we know that the $\sigma$ strings coincide with plaquette spin domain walls (at least for low energy states, which is all that we care about) so crossing a $\sigma$ string means moving from a domain where $\tau^z=+(-)$ to one where $\tau^z=-(+)$. We can therefore choose a convention in which one component, say $W_1$, is in the $\psi$ channel in the $\tau^z=+$ domain and $W_2$ is in the $\psi$ channel in the $\tau^z=-$ domain. Labeling these two components by $W_{\psi+}$ and $W_{\psi-}$ respectively, we can write
\beq\label{e+m}
W_{\sigma\bar{\sigma}} =W_{\psi+}+W_{\psi-}
\eeq
%which makes explicit reference to the $\tau$ spins. 

We now repeat this analysis for the second operator, $W_{\sigma\bar{\sigma}\bot\psi\bar{\psi}}$. However, before we can do that, we need to discuss $W_{\psi \bar{\psi}}$. This operator can be described using the same graphical representation as $W_{\sigma \bar{\sigma}}$. The only difference is that in this case, the first rule is replaced by
\begin{align}
  & \left|
  \begin{minipage}[h]{0.15\linewidth}
        \vspace{0pt}
        \includegraphics[height=.3in]{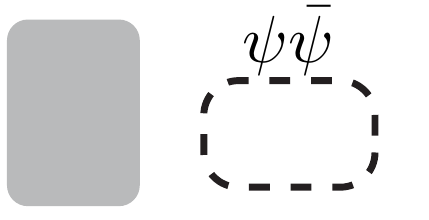}
   \end{minipage}\right\rangle=
 \left|
  \begin{minipage}[h]{0.15\linewidth}
        \vspace{0pt}
        \includegraphics[height=.3in]{resolved-loop-1.pdf}
   \end{minipage}\right\rangle\ 
\end{align}
and the $\Omega, \bar{\Omega}$ tensors are different. In particular, the only nonzero values of $\Omega^j_{sti} = \bar{\Omega}^j_{sti}$ are
\begin{equation}
\Omega^0_{000}=\Omega^2_{002}= 1, \ \ \Omega^1_{001} = -1
\end{equation}
in this case. Translating this graphical definition into algebra, one can see that the action of $W_{\psi \bar{\psi}}$ is very simple:
\beq \label{ntimes}
W_{\psi\bar{\psi}}=(-1)^{N_{\sigma\times}}
\eeq
where $N_{\sigma\times}$ is an operator that is diagonal in the string-net basis and that counts the total number of $\sigma$ strings that cross $\gamma$.

With this preparation, we are now ready to discuss the operator $W_{\sigma\bar{\sigma}\bot\psi\bar{\psi}}$. In the graphical representation, this operator inserts a $T$-junction made up of $\sigma \bar{\sigma}$ and $\psi \bar{\psi}$ strings. Crossings are then resolved using the $\sigma \bar{\sigma}$ and $\psi \bar{\psi}$ rules listed above. Translating this graphical definition into the notation of (\ref{w1w2}) and (\ref{ntimes}), one finds:
\beq \label{sigmapsi}
W_{\sigma\bar{\sigma}\bot\psi\bar{\psi}}=(W_1-W_2)(-1)^{N_{\sigma\times}}\eta_p
\eeq
where $\eta_p = \pm 1$ depending on whether $W_1$ is in the $\psi$ channel or $1$ channel at the point $p$ where $\psi\bar{\psi}$ attaches to $\sigma\bar{\sigma}$ (Fig. \ref{e-m-string}).

Now consider ungauging $W_{\sigma\bar{\sigma}\bot\psi\bar{\psi}}$ --- that is, embedding this operator in the Hilbert space of the symmetric toric code. In this new context, we can rewrite $(-1)^{N_{\sigma \times}}$ as
\begin{equation}\label{tautau}
(-1)^{N_{\sigma\times}} = \tau_p^z \tau_q^z
\end{equation}
since the $\sigma$ strings coincide with the plaquette spin domain walls at low energies (Fig. \ref{e-m-string}). We can also rewrite
$(W_1-W_2) \eta_p$ as:
\begin{equation}\label{w1-w2}
(W_1 - W_2) \eta_p = (W_{\psi+} - W_{\psi-}) \tau_p^z
\end{equation}
Multiplying (\ref{tautau}) and (\ref{w1-w2}) together and using (\ref{sigmapsi}), we derive
\begin{align}
\label{e--m}
W_{\sigma\bar{\sigma}\bot\psi\bar{\psi}}=(W_{\psi+}-W_{\psi-})\tau_q^z
\end{align}

We now have everything we need to compute $W_e$ and $W_m$. Combining equations (\ref{eplusm}, \ref{eminusm}, \ref{e+m}, \ref{e--m}) we derive:
\begin{align}\label{string-ops-ungauging}
&W_e\sim W_{\psi+}, \quad W_m\sim W_{\psi-}
\end{align}
As a final step, we multiply the string operators by the projector $P_\gamma$ from (\ref{string-projector}) which projects onto low energy states where $\sigma$ strings are bound to domain walls and \ {fusion} rules are obeyed. While this step is not strictly necessary, it ensures that the string operators are especially well-behaved. After this step, we have:
\begin{align}\label{string-ops-ungauging}
&W_e^{\gamma}= \mathcal{P}_{\gamma}W_{\psi+}\mathcal{P}_{\gamma} \nonumber \\
&W_m^{\gamma}= \mathcal{P}_{\gamma}W_{\psi-}\mathcal{P}_{\gamma}
\end{align}
To go from the above expressions (\ref{string-ops-ungauging}) to the formulas (\ref{w_e}) and (\ref{w_m}) in the main text is simply a matter of conversion from the graphical representation to an algebraic representation. In particular the $(-1)$ factors associated with the first $3$ vertices in Fig. \ref{vertices} ($e1$, $e2$ and $e3$) arise from fusing the strings created by $W_{\psi+}$ onto the honeycomb lattice. Likewise, the $(-1)$ factors from the last three vertices ($e4$, $e5$ and $e6$) come from the fact that $\Omega^2_{002}=-1$ for the $\sigma \bar{\sigma}$ string.

\end{appendix}

\bibliography{mybib}%{}
%\bibliographystyle{plain}
%%%%%%%%%%%%%%%%%%%%%%%%%%%%%%%%
%%%%%%%%%%%%%%%%%%%%%%%%%%%%%%%%

\end{document}